\documentclass{SciPost}

\hypersetup{
    colorlinks,
    linkcolor={red!50!black},
    citecolor={blue!50!black},
    urlcolor={blue!80!black}
}

\usepackage[bitstream-charter]{mathdesign}
\DeclareSymbolFont{usualmathcal}{OMS}{cmsy}{m}{n}
\DeclareSymbolFontAlphabet{\mathcal}{usualmathcal}

\fancypagestyle{SPstyle}{
\fancyhf{}
\lhead{\colorbox{scipostblue}{\bf \color{white} ~SciPost Physics }}
\rhead{{\bf \color{scipostdeepblue} ~Submission }}

\fancyfoot[C]{\textbf{\thepage}}
}

\usepackage{bm}
\usepackage{braket}
\usepackage{amsthm}
\newtheorem{theorem}{Theorem}

\newtheorem{lemma}{Lemma}
\newtheorem{criterion}{Criterion}

\usepackage{subcaption}
\begin{document}

\pagestyle{SPstyle}

\begin{center}{\Large \textbf{\color{scipostdeepblue}{
Numerical Block Diagonalization and Linked-Cluster Expansion for Deriving Effective Hamiltonians: Applications to Spin Excitations\\
}}}\end{center}

\begin{center}\textbf{
Tsutomu Momoi\textsuperscript{1$\star$} and
Owen Benton\textsuperscript{2}
}\end{center}

\begin{center}
{\bf 1} RIKEN Center for Emergent Matter Science, Wako, Saitama, 351-0198, Japan
\\
{\bf 2} School of Physical and Chemical Sciences, Queen Mary University of London, London, E1 4NS, United Kingdom
\\[\baselineskip]
$\star$ \href{mailto:email1}{\small momoi@riken.jp}
\end{center}

\section*{\color{scipostdeepblue}{Abstract}}
\textbf{\boldmath{%
We present a numerical, non-perturbative framework for constructing effective Hamiltonians
that describe the dynamics of low-energy degrees of freedom within a restricted Hilbert space
in quantum many-body systems.
The approach is based on block diagonalization guided by a minimal-deformation principle
imposed within a selected target sector.
The formulation is designed to remain compatible with the numerical linked-cluster expansion.
For gapped systems, the relation between minimal deformation and cluster additivity
requires careful treatment when excited eigenstates contain finite admixtures of the ground state.
After establishing a cluster-additive basis that reproduces the H\"ormann–Schmidt construction,
the minimal-deformation criterion uniquely determines the effective Hamiltonian within each excitation sector.
The same criterion also provides a practical numerical procedure
for selecting relevant low-energy eigenstates,
including regimes characterized by strong level mixing and avoided crossings.
The framework is illustrated using two spin models:
the one-dimensional transverse-field Ising model as a benchmark
and the two-dimensional Shastry–Sutherland model with Dzyaloshinskii–Moriya interactions,
relevant to SrCu$_2$(BO$_3$)$_2$.
In both cases, the resulting effective Hamiltonians accurately capture the excitation dynamics
and the associated band structures.
}}

\vspace{\baselineskip}

\noindent\textcolor{white!90!black}{%
\fbox{\parbox{0.975\linewidth}{%
\textcolor{white!40!black}{\begin{tabular}{lr}%
  \begin{minipage}{0.6\textwidth}%
    {\small Copyright attribution to authors. \newline
    This work is a submission to SciPost Physics. \newline
    License information to appear upon publication. \newline
    Publication information to appear upon publication.}
  \end{minipage} & \begin{minipage}{0.4\textwidth}
    {\small Received Date \newline Accepted Date \newline Published Date}%
  \end{minipage}
\end{tabular}}
}}
}


\vspace{10pt}
\noindent\rule{\textwidth}{1pt}
\tableofcontents
\noindent\rule{\textwidth}{1pt}
\vspace{10pt}


\section{Introduction}

In quantum many-body systems, low-energy properties are often governed by a restricted subspace of the full Hilbert space.
Effective Hamiltonians provide an effective description within a chosen target subspace
while incorporating the influence of high-energy degrees of freedom.
A common strategy is to construct such Hamiltonians through perturbative or non-perturbative transformations that decouple the low- and high-energy sectors~\cite{VanVleck1929,Bloch1958,Schrieffer1966,GlazekWilson1993,GlazekWilson1994,Wegner1994,Mielke1998,Kehrein2006}.

Traditional approaches frequently rely on perturbative expansions, which generate analytic effective Hamiltonians order by order in a small control parameter~\cite{desCloizeaux1960,Takahashi1977,Shavitt1980,Durand1983,oitmaa_hamer_zheng_2006,Knetter2000A,Knetter2003}.
Although these constructions are conceptually transparent, they are limited to regimes where the expansion remains controlled.
In many systems of current interest, however, the relevant couplings are not perturbatively small, and
strong level mixing invalidates purely perturbative treatments.

Several non-perturbative frameworks have therefore been developed.
A prominent example is the continuous unitary transformation (CUT) method~\cite{Wegner1994,Mielke1998,Kehrein2006},
which achieves block diagonalization through flow equations without requiring a small expansion parameter.
More recently, hybrid approaches~\cite{YangS2011,YangACLS2012,Ixert2014,Coester2015} combining CUT with the numerical linked-cluster
expansion (NLCE)~\cite{RigolBS2006,RigolBS2007,Tang2013}
have enabled the construction of effective models on finite clusters, followed by non-perturbative extrapolation to the thermodynamic limit.
Related projective block-diagonalization schemes combined with NLCE have also been proposed~\cite{Hormann2023}.
These developments motivate a unified and numerically robust framework for constructing effective Hamiltonians beyond perturbation theory.

A noteworthy feature of block-diagonalization schemes is their non-uniqueness:
many unitary or similarity transformations can decouple the same target subspace
while yielding different effective Hamiltonians.
A natural guiding principle is therefore to select the transformation
that modifies the target basis as little as possible~\cite{CederbaumSM}.
This minimal-deformation principle underlies several constructions,
including the Schrieffer--Wolff transformation~\cite{Schrieffer1966,BRAVYI2011},
projective approaches~\cite{desCloizeaux1960,Takahashi1977},
and the Cederbaum--Schirmer--Meyer (CSM) transformation~\cite{CederbaumSM}.

When effective Hamiltonians are combined with NLCE, an additional requirement arises:
\emph{cluster additivity}~\cite{Gelfand1996,Trebst2000,Knetter2003}.
For disconnected clusters $A$ and $B$,
the effective Hamiltonian must satisfy
\begin{equation}
{\bm H}_{\rm eff}^{A\cup B}
=
{\bm H}_{\rm eff}^{A}\otimes {\bm 1}^B
+
{\bm 1}^A \otimes {\bm H}_{\rm eff}^{B}.
\end{equation}
This property is automatically satisfied when the effective Hamiltonian is constructed
within a degenerate low-energy manifold~\cite{BRAVYI2011}.
However, it is not generically guaranteed for gapped systems with a unique ground state.
In such cases, applying minimal-deformation schemes directly to excitation sectors
may violate cluster additivity when excited eigenstates acquire finite admixtures of the ground
state~\cite{Hormann2023}.

H\"ormann and Schmidt (HS) restored cluster additivity in gapped systems through a state-level transformation
within sectors of fixed excitation number~\cite{Hormann2023}.
They further proposed a cluster-additive effective-Hamiltonian construction that generalizes the projective transformation.
While the resulting effective-Hamiltonian construction was introduced as a generalized minimal transformation, it is not immediately transparent how this generalized construction realizes the minimal-deformation principle in the CSM sense. 
In the present work, we show that the HS transformation can be reformulated as defining a cluster-additive basis
on which a minimal-deformation criterion can subsequently be imposed.

In this work, we formulate a numerical block-diagonalization framework that makes explicit the
relation between minimal deformation and cluster additivity and is fully compatible with NLCE.
The key idea is to formulate the minimal-deformation principle
directly within the chosen target sector rather than in terms of the full Hilbert-space transformation.
This sector-restricted formulation applies both to systems with quasi-degenerate low-energy manifolds and
to gapped systems with a unique ground state.
For gapped systems, since minimal deformation alone does not guarantee cluster additivity,
we first construct a cluster-additive basis that reproduces the HS state transformation and
subsequently impose a minimal-deformation criterion within this basis.
This two-step procedure yields exactly the same effective Hamiltonian as the HS construction,
while providing an explicit minimal-deformation interpretation within a cluster-additive basis.
The same minimization principle also provides a practical criterion for selecting relevant eigenstates
in the presence of avoided level crossings and strong mixing induced by particle-number-nonconserving interactions.
When combined with NLCE, the framework yields a numerically robust construction of effective Hamiltonians
in the thermodynamic limit.

To illustrate the scope and performance of the method, we apply it to two spin models.
As a benchmark, we consider the one-dimensional transverse-field Ising model, which is exactly solvable~\cite{Pfeuty1970},
and compare our results with the exact one-magnon dispersion.
As a realistic application, we study the two-dimensional Shastry--Sutherland model~\cite{shastry1981}
with Dzyaloshinskii--Moriya (DM) interactions, relevant to the spin-dimer compound
SrCu$_2$(BO$_3$)$_2$~\cite{Kageyama1999,Miyahara1999}.
In this system, the low-field phase hosts gapped triplon excitations whose weak dispersion
is induced by DM couplings~\cite{Cepas2001,Nojiri2003,Room2004,Gaulin2004,Gozar2005,Cheng2007,Romhanyi2011,Romhanyi2015,Malki2017,McClarty2017},
and, due to a relatively large perturbative parameter
$J^\prime/J\simeq 0.6$--0.63~\cite{Miyahara2000,Knetter2000B,Marurenko2008},
perturbative treatments are not guaranteed to remain quantitatively reliable for realistic parameters~\cite{Weihong1999,Koga2000,Momoi2000A,Momoi2000,Fukumoto2000,Knetter2000B,Totsuka2001,Dorier2008}.
Our non-perturbative framework captures the resulting triplon dynamics and the emergence of topological band structures.


The remainder of this paper is organized as follows.
In Sec.~2, we present the numerical block-diagonalization framework and formulate the minimal-deformation criteria that
uniquely determine the effective Hamiltonian in each target sector.
Sections~\ref{sec:1d} and \ref{sec:2d} present applications to the one-dimensional transverse-field Ising model
and the two-dimensional Shastry--Sutherland model with DM interactions, respectively.
We conclude with a summary and outlook in Sec.~\ref{sec:summary}, while technical details are deferred to the appendices.

\section{Method: construction of effective Hamiltonians}\label{sec:method}

This section presents a numerical block-diagonalization scheme
based on a sector-wise minimal-deformation principle
and its implementation within NLCE.
Section~\ref{sec:framework} introduces the general framework. Section~\ref{sec:gap} discusses gapped systems
and cluster additivity. Section~\ref{sec:select} presents the eigenstate-selection procedure.
Secstions~\ref{sec:nlc}–\ref{sec:asympt} describe the NLCE construction and the asymptotic structure
of the resulting effective Hamiltonians.


\subsection{General framework}\label{sec:framework}

As emphasized in the Introduction, block diagonalization of a many-body Hamiltonian is not unique:
many different unitary or similarity transformations can decouple the same target subspace
while producing distinct effective Hamiltonians.
Here we formulate a general framework and introduce a minimal-deformation criterion that uniquely fixes the effective Hamiltonian
within a chosen target sector, using only low-energy eigenvalues and eigenvectors.

\subsubsection{Notation and block structure}\label{sec:notation}

We fix notation for eigenstates, eigenvector matrices, and the block decomposition of the Hilbert space,
associated with a chosen set of quantum numbers that defines the sector structure.

We consider a Hamiltonian of the form
\begin{equation}
H = H_0 + V,
\end{equation}
where $H_0$ denotes the unperturbed Hamiltonian and $V$ represents a perturbation.
The Hamiltonian $H_0$ is assumed to define a sector decomposition through a conserved quantum number (or an
excitation number),
which serves as a natural reference block structure.
Typical examples include the total $S^z$ in polarized magnets and the triplon number
in dimerized systems.
Let $\{ |a_i\rangle \}$ and $\{ |\phi_i\rangle \}$ ($i=1,\dots,N$) denote the orthonormal eigenstates of $H_0$ and
the full Hamiltonian $H$, respectively.
The two sets of eigenstates are related through the eigenvector matrix ${\bm S}$ as
\begin{equation}\label{eq:def_S_method}
\left( |\phi_1\rangle, |\phi_2\rangle, \dots, |\phi_N\rangle \right)
= \left( |a_1\rangle, |a_2\rangle, \dots, |a_N\rangle \right) {\bm S},
\end{equation}
where ${\bm S}$ satisfies the unitarity condition ${\bm S}^\dagger {\bm S} = {\bm 1}$.

In the basis of $\{ |a_i\rangle \}$, the full Hamiltonian matrix ${\bm H}$ is given by
$[{\bm H}]_{ij} = \langle a_i| H | a_j \rangle$.
The eigenvector matrix ${\bm S}$ then diagonalizes ${\bm H}$ as
\begin{align}\label{eq:full_diag}
  {\bm S}^\dagger {\bm H}{\bm S}= {\bm \Lambda}, \qquad  {\bm \Lambda}= \mathrm{diag}
    (\lambda_{1}, \lambda_2, \dots, \lambda_N ),
\end{align}
where $\lambda_i$ denote the eigenvalues of ${\bm H}$.

For later convenience, we introduce a block decomposition of the Hilbert space into subspaces
spanned by eigenstates of $H_0$ with different quantum numbers.
When the Hilbert space is partitioned into $m$ subspaces,
the Hamiltonian matrix ${\bm H}$ takes the block form
\begin{equation}
{\bm H} =
\begin{bmatrix}
{\bm H}_{11} & {\bm H}_{12} & \cdots & {\bm H}_{1m} \\
{\bm H}_{21} & {\bm H}_{22} & \cdots & {\bm H}_{2m} \\
\vdots & \vdots & \ddots & \vdots \\
{\bm H}_{m1} & {\bm H}_{m2} & \cdots & {\bm H}_{mm}
\end{bmatrix},
\end{equation}
where the off-diagonal blocks ${\bm H}_{ij}$ ($i\neq j$) arise from the perturbation $V$.
The goal of the block-diagonalization procedure discussed in Secs.~\ref{sec:BD_eff_H} and
\ref{sec:min_def_cluster_add} is to find a minimal block-diagonalizing transformation ${\bm T}$
that eliminates these off-diagonal couplings and yields a physically meaningful effective Hamiltonian.

In this context, we define the block structure of the eigenvector matrix $\bm S$ with respect to the reference basis
decomposed into sectors labeled by a quantum number.
Since the full Hamiltonian does not generally conserve this quantum number,
its exact eigenstates are superpositions of different sectors.
We therefore assign eigenstates to a given sector by selecting those with dominant weight in the corresponding subspace.
This assignment is further guided by the minimal-deformation criterion introduced below,
which selects states that remain as close as possible to the reference subspace.

\subsubsection{Minimal block diagonalization and a sector-wise criterion}\label{sec:BD_eff_H}

We briefly review the minimal block-diagonalization scheme of Cederbaum, Schirmer, and Meyer (CSM)~\cite{CederbaumSM}
and introduce a sector-wise minimality criterion that directly targets the low-energy subspace of interest.
This formulation allows the effective Hamiltonian in the target subspace to be constructed from low-energy
eigenstate information only.

Block diagonalization of the Hermitian matrix ${\bm H}$ is performed by a unitary matrix ${\bm T}$ such that
\begin{align}\label{eq:block_d}
{\bm T}^{\dagger} {\bm H} {\bm T} =
  \begin{bmatrix}
    {\bm H}_{{\rm eff},11} &   &   & 0 \\
      & {\bm H}_{{\rm eff},22} &   &   \\
      &   &\ddots &   \\
    0 &   &   & {\bm H}_{{\rm eff},mm}
  \end{bmatrix},
\end{align}
where each ${\bm H}_{{\rm eff},ii}$ is an $n_i \times n_i$ Hermitian matrix, and
all off-diagonal blocks vanish.
The transformed Hamiltonian corresponds to the matrix representation of  $H$ in the new basis
$|b_i\rangle=\sum_{j=1}^N |a_j\rangle [{\bm T}]_{ji}$
$(i=1,\cdots, N)$, i.e.,
$\langle b_i|H|b_j\rangle=[{\bm T}^{\dagger} {\bm H} {\bm T}]_{ij}$.

Since ${\bm S}$ diagonalizes ${\bm H}$ as
${\bm S}^{\dagger}{\bm H}{\bm S}={\bm \Lambda}$,
any unitary transformation that achieves this block diagonalization
can be written in the form
\begin{equation}\label{eq:SF}
{\bm T} = {\bm S}{\bm F},
\end{equation}
where
${\bm F}$ is an arbitrary block-diagonal unitary matrix
consistent with the chosen subspace decomposition.
This yields
${\bm T}^{\dagger} {\bm H} {\bm T} = {\bm F}^{\dagger} {\bm \Lambda} {\bm F}$, which is block-diagonal.
Since ${\bm F}$ is not unique, the transformation matrix ${\bm T}$ is not unique.

To remove this arbitrariness, CSM~\cite{CederbaumSM} required the transformation to
be minimal.
This condition selects, among all block-diagonalizing transformations,
the one closest to the identity.
Specifically, they imposed the following minimal-deformation condition in the Frobenius norm:
\begin{equation}\label{eq:condition}
\| {\bm T} -{\bm 1}_N \|_F= {\rm minimum},
\end{equation}
where ${\bm 1}_N$ denotes the $N\times N$ identity matrix.
This norm measures the distance between the transformed basis and
the original one.
Under this condition, the optimal transformation is uniquely given by
${\bm F}={\bm S}_{\rm BD}^\dagger({\bm S}_{\rm BD}{\bm S}_{\rm BD}^\dagger)^{-1/2}$, i.e.,
\begin{align}\label{eq:unitry_T}
  {\bm T}= {\bm S}{\bm S}_{\rm BD}^\dagger({\bm S}_{\rm BD}{\bm S}_{\rm BD}^\dagger)^{-1/2},
\end{align}
where
${\bm S}_{\rm BD}$ is the block-diagonal part of ${\bm S}$ with respect to the chosen subspace decomposition,
\begin{align}\label{eq:S}
{\bm S}_{\rm BD}=
  \begin{bmatrix}
    {\bm S}_{11} &  &  & 0 \\
     & {\bm S}_{22} &  &  \\
     &  &\ddots &  \\
    0 &  &  & {\bm S}_{mm} \\
  \end{bmatrix}.
\end{align}
This transformation is equivalent to the projective transformation~\cite{desCloizeaux1960,Takahashi1977}
and to the Schrieffer--Wolff transformation~\cite{Schrieffer1966,BRAVYI2011},
both originally introduced in perturbation theory.
See, for example, Ref.~\cite{Hormann2023} for a comprehensive review.

\paragraph{Sector-wise minimality criterion}

To make the framework more suitable for practical construction of low-energy effective Hamiltonians,
we introduce a sector-wise minimality criterion that extends the CSM construction to a target subspace.
Let the target low-energy subspace correspond to the first sector of the partitioned subspaces,
with dimension $n$.
The corresponding block ${\bm T}_{11}$ of the unitary matrix ${\bm T}$ acts within this subspace.
To ensure minimal deformation within the target subspace,
we impose the following criterion.

\begin{criterion}
Among all unitary transformations ${\bm T}$ that block-diagonalize ${\bm H}$,
we choose ${\bm T}$ such that its restriction to the target subspace satisfies
the minimality condition
\begin{align}\label{eq:min_condition}
  \| {\bm T}_{11} -{\bm 1}_n \|_F= {\rm minimum}.
\end{align}
\end{criterion}

This quantity has a clear geometric meaning.
The matrix ${\bm T}$ transforms the original basis $\{\ket{a_i}\}$ to a new basis $\{\ket{b_i}\}$.
The block ${\bm T}_{11}$ describes the overlap between the original
basis of the target subspace and the transformed basis projected back onto the same subspace.
Thus, $\|{\bm T}_{11}-{\bm 1}_n\|$
measures the distance between theses bases, and
minimizing it selects the block-diagonalizing transformation
that minimally distorts the target subspace.
This criterion, together with the analogous minimal-deformation condition introduced in
Sec.~\ref{sec:min_def_cluster_add}, will serve as a guiding principle
for constructing effective Hamiltonians.

\paragraph{Effective Hamiltonian}

To derive an explicit expression for ${\bm T}_{11}$,
we use the singular value decomposition (SVD) of ${\bm S}_{11}$,
which is given by
\begin{align}
{\bm S}_{11} = {\bm U}_1 {\bm \Sigma}_1 {\bm V}_1^{\dagger}.
\label{eq:SVD}
\end{align}
Here ${\bm U}_1$ and ${\bm V}_1$ are $n\times n$ unitary matrices and
${\bm \Sigma}_1$ is an $n\times n$ diagonal matrix with non-negative entries.
Using this decomposition, the minimal transformation ${\bm T}_{11}$
and the corresponding effective Hamiltonian can be obtained explicitly,
as stated in the following theorem.

\begin{theorem}\label{thm:1}
Assume that $\bm{S}_{11}$ is invertible.
Then the transformation $\bm{T}=\bm{S}\bm{F}$ satisfying Criterion 1
is uniquely determined in the first block by
\begin{align}\label{eq:T11}
{\bm T}_{11} & = {\bm U}_1 {\bm \Sigma}_1 {\bm U}_1^\dagger
\end{align}
and the corresponding effective Hamiltonian block is
\begin{align}\label{eq:eff_H11}
{\bm H}_{{\rm eff},11}
&= {\bm U}_1 {\bm V}_1^\dagger {\bm \Lambda}_1 {\bm V}_1 {\bm U}_1^\dagger,
\end{align}
where ${\bm \Lambda}_1$ is the diagonal block of ${\bm \Lambda}$ in the 1st sector.
\end{theorem}

The proof is given in Appendix~\ref{sec:proof}.
The resulting effective Hamiltonian in Eq.~(\ref{eq:eff_H11}) coincides with the effective Hamiltonian
obtained from the CSM construction.

Although the proof closely follows the original CSM argument,
the resulting formulation is advantageous for numerical applications.
Once the low-energy eigenstates and eigenvalues are obtained, the block ${\bm T}_{11}$ and
the effective Hamiltonian ${\bm H}_{{\rm eff},11}$ can be constructed using only information within this sector.
This formulation also enables the minimal-deformation criterion to be used for the
eigenstate-selection procedure described in Sec.~\ref{sec:select}.
Accordingly, the present framework can be implemented using only low-energy data,
without requiring access to high-energy eigenstates.

Before proceeding, it is useful to clarify under which conditions
the above construction remains compatible with cluster additivity.
When the target space forms a degenerate low-energy manifold,
the block-diagonalization procedure automatically preserves
cluster additivity~\cite{BRAVYI2011,Hormann2023}, because disconnected clusters factorize
within the manifold.
In contrast, for systems with a unique ground state,
excited eigenstates may contain finite admixtures of the ground state.
In such cases, the block structure defined by excitation sectors
does not factorize on disconnected clusters,
and cluster additivity can be violated~\cite{Hormann2023}.
This distinction motivates the specialized construction
presented in the following subsection.

\subsection{Gapped systems: cluster additivity}\label{sec:gap}

Building on the general framework introduced in Sec.~\ref{sec:framework},
we now consider gapped systems with a unique ground state,
for which cluster additivity is not automatically guaranteed
within the minimal-deformation framework.
The difficulty arises when excited eigenstates contain finite admixtures of the
unperturbed ground state, i.e., when the target excitation manifold is directly
coupled to the ground-state sector.
This situation typically occurs when the perturbation $V$ does not conserve the excitation number
defined by $H_0$.

To address this issue, we separate the construction into two steps.
First, we construct a cluster-additive basis that removes ground-state admixtures
from the excitation sectors, following Ref.~\cite{Hormann2023}.
Second, within this basis, we impose
the minimal-deformation criterion on the block-diagonalizing transformation.

\subsubsection{Irreducible effective Hamiltonians}\label{sec:irreducible}

To enforce cluster additivity for excitation energies,
we organize the Hilbert space into excitation-number sectors and introduce irreducible effective Hamiltonians.
This is achieved by recursively subtracting all lower-excitation contributions from the effective Hamiltonian
defined in a fixed excitation-number sector~\cite{Gelfand1996,Trebst2000,Knetter2003}.
This step enforces additivity of excitation energies, but, as we explain later, does not fully ensure cluster additivity.

To this end, we decompose the Hilbert space on a cluster $C$ into excitation-number sectors,
\begin{equation}
\mathcal{H}_C = \bigoplus_{n=0}^{n_{\max}} \mathcal{H}_C^{(n)},
\end{equation}
where $\mathcal{H}_C^{(n)}$ denotes the subspace containing $n$ excitations and
$n_{\max}$ is the maximum number of excitations allowed.

To ensure cluster additivity of excitation energies,
it is necessary to remove contributions originating from lower excitation sectors.
This is achieved by introducing the irreducible $n$-excitation effective Hamiltonian
$\bar{{H}}^{(n)}_{\rm eff}$,
defined recursively by subtracting all lower-particle contributions
from the effective Hamiltonian obtained in the $n$-excitation sector~\cite{Gelfand1996,Trebst2000,Knetter2003}:
\begin{align}\label{eq:subtract}
  \bar{{H}}^{(0)}_{\rm eff} & = {H}^{(0)}_{\rm eff}, \nonumber\\
  \bar{{H}}^{(n)}_{\rm eff} & =
  {H}^{(n)}_{\rm eff}
  - \sum_{i=0}^{n-1} \bar{{H}}^{(i)}_{\rm eff},
  \quad\quad (n>0).
\end{align}
Here ${H}^{(n)}_{\rm eff}$ denotes the effective Hamiltonian obtained
by block diagonalization within the $n$-excitation sector.
The lower-sector Hamiltonians act independently on subsets of excitations
once embedded into the $n$-particle sector.
The subtraction removes contributions associated with
independent propagation of fewer particles, so that
$\bar{{H}}^{(n)}_{\rm eff}$ contains only
irreducible $n$-excitation processes.

\subsubsection{Violation of cluster additivity}

Even after constructing irreducible effective Hamiltonians,
cluster additivity may still be violated in gapped systems
when excited eigenstates contain admixtures of the ground state.
Such admixtures arise when the perturbation does not conserve
the excitation number defined by the unperturbed Hamiltonian.
The mechanism can be understood by considering disconnected clusters,
as illustrated below in the one-excitation sector.

\paragraph{Notation.}
We denote by $\bm{S}^C$ the eigenvector matrix for the cluster $C$.
With respect to the excitation-number decomposition of $\mathcal{H}_C$,
$\bm{S}^C$ can be written in block form as $\bm{S}_{ij}$,
where the block $(i,j)$ maps the $(j-1)$-excitation sector to the $(i-1)$-excitation sector.

For index sets $I,J\subset\{1,\dots,n_{\max}+1\}$,
$\bm{S}_{I,J}$ denotes the corresponding submatrix.
We further define $I_k := \{1,2,\dots,k\}$, $I_{0} = \varnothing$,
and its complement $I_k^{\mathrm{c}} := \{k+1,\dots,n_{\max}+1\}$.
Subspaces labeled by ``$\ge m$'' in excitation number correspond to the index set
$I_{m-1}^{\mathrm c}$.

Throughout this work, \emph{cluster additivity} refers strictly to a property
of effective Hamiltonians defined on disconnected clusters.
For brevity, we call a basis (or eigenvector matrix)
\emph{cluster-additive} if, when expressed in the excitation-number (particle-number) decomposition,
its diagonal blocks are block-diagonal with respect to disconnected clusters,
which is sufficient to ensure cluster additivity of the resulting effective Hamiltonian.

\paragraph{Excitation-number sectors on disconnected clusters.}
For two disconnected clusters $A$ and $B$,
the Hilbert space factorizes as $\mathcal{H}_{A \cup B} = \mathcal{H}_A \otimes \mathcal{H}_B$.
The fixed excitation-number subspace then decomposes as
\begin{align}
\mathcal{H}_{A \cup B}^{(m)}
&=
\bigoplus_{n=0}^{m}
\left(
\mathcal{H}_A^{(m-n)} \otimes \mathcal{H}_B^{(n)}
\right).
\end{align}

\paragraph{Breakdown in the one-excitation sector.}
We illustrate this mechanism in the one-excitation sector $\mathcal{H}_{A \cup B}^{(1)}$.
With respect to the decomposition
\begin{equation}\label{eq:H_decom_one}
\mathcal{H}_{A \cup B}^{(1)}=
(\mathcal{H}_A^{(1)} \otimes \mathcal{H}_B^{(0)}) \oplus (\mathcal{H}_A^{(0)} \otimes \mathcal{H}_B^{(1)}),
\end{equation}
cluster additivity requires the eigenvector matrix in this sector to be block-diagonal
(see Ref.~\cite{Hormann2023}).

In the absence of ground-state admixtures, eigenvectors on
disconnected clusters factorize, leading to a block diagonal form
\begin{equation}
\bm{S}_{22}^{A \cup B}
=
\left(\bm{S}_{22}^A \otimes S_{11}^B\right)
\oplus
\left(S_{11}^A \otimes \bm{S}_{22}^B\right).
\end{equation}
Here $S_{11}^{A}$ and $S_{11}^{B}$ denote the ground-state
components of the eigenvector matrices on clusters $A$ and $B$.
However, if the excited eigenvectors contain a finite overlap with the
unperturbed ground state, this factorization breaks down and the
$(2,2)$ block of $\bm S^{A\cup B}$ is no longer block-diagonal,
\[
\bm{S}_{22}^{A \cup B}
\neq
\left(\bm{S}_{22}^A \otimes S_{11}^B\right)
\oplus
\left(S_{11}^A \otimes \bm{S}_{22}^B\right).
\]
This mixing prevents the excitation sector from factorizing
into independent contributions on disconnected clusters.

Importantly, this difficulty cannot be resolved by modifying
the minimal-deformation condition alone.
The breakdown arises because the eigenvector
matrix itself does not factorize with respect to disconnected
clusters when excited states contain ground-state admixtures.
Consequently, minimal-deformation block-diagonalization
performed in the original basis does not, in general,
guarantee cluster additivity of the resulting effective
Hamiltonian.
This observation shows that cluster additivity must
first be restored at the level of the basis before a
minimal block-diagonalization criterion can be meaningfully applied.

\subsubsection{Restoration via the H{\"o}rmann--Schmidt transformation}
\label{subsec:cluster_additivity_restoration}

To restore cluster additivity, we construct a basis
in which excitation sectors factorize on disconnected clusters.
A transformation that achieves this was proposed by H{\"o}rmann and Schmidt (HS),
who introduced a state-level construction that restores cluster additivity
within fixed excitation-number sectors~\cite{Hormann2023}.
This transformation removes the overlap between excited eigenvectors and the unperturbed ground state,
ensuring that the relevant diagonal blocks become block-diagonal for disconnected clusters.
Below we summarize the construction in the one-excitation sector and then describe its generalization to the $m$-excitation sector.

\paragraph{One-excitation sector.}

In the one-excitation sector,
HS introduced a state transformation~\cite{Hormann2023}
that eliminates the overlap between one-excitation eigenvectors and the unperturbed ground state.
With the decomposition
$\mathcal{H}=\mathcal{H}^{(0)}\oplus\mathcal{H}^{(1)}\oplus\mathcal{H}^{(\ge 2)}$,
the transformed eigenvector matrix reads
\begin{equation}\label{eq:HS_transform}
\bar{\bm{S}} =
\begin{bmatrix}
S_{11} & \bm{0} & \bm{S}_{1,I_2^{\mathrm{c}}}
\\[3pt]
\bm{S}_{21} & \bm{S}_{22}- S_{11}^{-1}\bm{S}_{21}\bm{S}_{12}
& \bm{S}_{2,I_2^{\mathrm{c}}}
\\[3pt]
\bm{S}_{I_2^{\mathrm{c}},1} &
\bm{S}_{I_2^{\mathrm{c}},2}- S_{11}^{-1}\bm{S}_{I_2^{\mathrm{c}},1}\bm{S}_{12}
& \bm{S}_{I_2^{\mathrm{c}},I_2^{\mathrm{c}}}
\end{bmatrix}.
\end{equation}
For two disconnected clusters $A$ and $B$, the transformed $(2,2)$ block becomes block-diagonal with
respect to this decomposition,
\begin{align}
\bar{\bm{S}}_{22}^{A \cup B}
&=
\bar{\bm{S}}_{22}^A \bar{S}_{11}^B
\oplus
\bar{\bm{S}}_{22}^B \bar{S}_{11}^A ,
\end{align}
with respect to the decomposition \eqref{eq:H_decom_one}.
Within the projection-based construction of Ref.~\cite{Hormann2023},
this block-diagonal structure is sufficient to restore cluster additivity
of the one-excitation effective Hamiltonian.
As we show below, this construction coincides
with the effective Hamiltonian derived from a minimal-deformation criterion.

\paragraph{$m$-excitation sectors.}

HS further proposed a state transformation for
general $m$-excitation sectors~\cite{Hormann2023}.
In their construction, the $(m+1,m+1)$ block is transformed as
\begin{equation}
\bar{\bm{S}}_{m+1,m+1}
=
\bm{S}_{m+1,m+1}
-
\bm{S}_{m+1,I_m}
\left(\bm{S}_{I_m,I_m}\right)^{-1}
\bm{S}_{I_m,m+1},
\label{eq:HS_m_particle}
\end{equation}
i.e., the Schur complement with respect to the
leading principal block $\bm{S}_{I_m,I_m}$.
It is not immediately evident from Ref.~\cite{Hormann2023} that
the resulting block $\bar{\bm{S}}_{m+1,m+1}$ is block-diagonal
for disconnected clusters;
a sketch of the proof is provided in Appendix~\ref{sec:proof_cluster_add}.

In the following subsections, the state transformations are reformulated
as similarity transformations of the basis, and
effective Hamiltonians are constructed from the resulting diagonal blocks using minimal transformations.
The resulting effective Hamiltonian coincides with that derived from the projection-based construction
of H\"ormann and Schmidt.

\subsubsection{Similarity basis transformations}
\label{subsec:similarity_basis_transformation}

The HS prescription restores cluster additivity at the level of eigenvectors and provides a projection-based
construction of effective Hamiltonians.
To clarify its relation to the minimal-deformation principle,
we reformulate the HS transformation as a similarity transformation of the basis.
This reformulation provides a natural setting to impose the minimal-deformation criterion directly
within a cluster-additive basis and offers a practical route for numerical implementation.
We first present the construction in the one-excitation sector and then develop an inductive scheme applicable to
the general $m$-excitation sector.

We introduce a new basis defined by
\begin{equation}
\left(\ket{\tilde{a}_1}, \ldots, \ket{\tilde{a}_N}\right)
=
\left(\ket{a_1}, \ldots, \ket{a_N}\right)\,\bm{W},
\label{eq:basis_sim_transf}
\end{equation}
through an invertible linear transformation $\bm{W}$.
In this new basis, the eigenvector matrix and Hamiltonian transform as
\begin{equation}
\tilde{\bm{S}} = \bm{W}^{-1}\bm{S},
\qquad
\tilde{\bm{H}} = \bm{W}^{-1}\bm{H}\bm{W}.
\label{eq:similarity_H}
\end{equation}

\paragraph{One-excitation sector}

From Ref.~\cite{Hormann2023}, cluster additivity
in the one-excitation effective Hamiltonian is restored if
\begin{equation}
\tilde{S}_{11} = S_{11}, \qquad
\tilde{\bm{S}}_{22} =
\bm{S}_{22} - S_{11}^{-1}\,\bm{S}_{21}\,\bm{S}_{12}.
\label{eq:HS_condition_22}
\end{equation}
Within the projection-based construction of Ref.~\cite{Hormann2023},
these conditions are sufficient to restore cluster additivity in the one-excitation effective Hamiltonian.
They are not necessary conditions, and alternative cluster-additive constructions may exist.

In the present approach we choose the similarity transformation
$\bm{W}$ so that Eq.~\eqref{eq:HS_condition_22} is satisfied.
This choice is not unique:
many similarity transformations reproduce the same diagonal blocks
$\tilde{S}_{11}$ and $\tilde{\bm{S}}_{22}$.
With respect to the decomposition
$\mathcal{H}=\mathcal{H}^{(0)}\oplus\mathcal{H}^{(1)}\oplus\mathcal{H}^{(\ge 2)}$,
we adopt the following convenient choice:
\begin{equation}\label{eq:w1}
\bm{W} =
\begin{bmatrix}
1 & \bm{0} & \bm{0} \\[3pt]
S_{11}^{-1}\,\bm{S}_{21} & {\bm 1} & \bm{0} \\[3pt]
S_{11}^{-1}\,\bm{S}_{I_2^\mathrm{c},1} & \bm{0} & {\bm 1}
\end{bmatrix}.
\end{equation}
The resulting similarity-transformed eigenvector matrix $\tilde{\bm{S}}$ is given by
\begin{equation}\label{eq:transformed_state_one}
\tilde{\bm{S}}= \bm{W}^{-1} \bm{S} =
\begin{bmatrix}
S_{11} & \bm{S}_{12} & \bm{S}_{1,I_2^\mathrm{c}} \\[3pt]
\bm{0} & \bm{S}_{22}- S_{11}^{-1} \bm{S}_{21} \bm{S}_{12}
& \bm{S}_{2,I_2^\mathrm{c}}- S_{11}^{-1} \bm{S}_{21} \bm{S}_{1,I_2^\mathrm{c}} \\[3pt]
\bm{0} & \bm{S}_{I_2^\mathrm{c},2}- S_{11}^{-1} \bm{S}_{I_2^\mathrm{c},1} \bm{S}_{12}
& \bm{S}_{I_2^\mathrm{c},I_2^\mathrm{c}}- S_{11}^{-1} \bm{S}_{I_2^\mathrm{c},1} \bm{S}_{1,I_2^\mathrm{c}}
\end{bmatrix}.
\end{equation}
Note that $\tilde{\bm{S}}$ does not coincide with
the HS state transformation $\bar{\bm{S}}$.
In general, the latter cannot be expressed purely as a similarity transformation of the basis.

\paragraph{$m$-excitation sector}

In the $m$-excitation sector,
we apply similarity transformations recursively:
\begin{equation}
\tilde{\bm{S}}^{(n)} = \bm{W}_n^{-1}\tilde{\bm{S}}^{(n-1)},
\qquad \tilde{\bm{S}}^{(0)} = \bm{S}.
\end{equation}
The first step $\bm{W}_1$ coincides with Eq.~\eqref{eq:w1}.
The total transformation is
\begin{equation}
\bm{W} \equiv \bm{W}_1 \bm{W}_2 \cdots \bm{W}_m
\label{eq:W_definition}
\end{equation}
and the transformed matrix is given by
\begin{equation}
\tilde{\bm{S}}^{(m)} = \bm{W}^{-1}\bm{S}
= \bm{W}_m^{-1}\cdots \bm{W}_1^{-1}\bm{S}.
\end{equation}

Each transformation $\bm{W}_n$
acts nontrivially only on sectors with excitation number greater than or equal to $n-1$.
Its role is to remove the coupling between the $(n-1)$-excitation sector and
higher sectors by a Schur-complement reduction
with respect to the $(n,n)$ block of $\tilde{\bm{S}}^{(n-1)}$.
To define $\bm{W}_n$ explicitly, we introduce the decomposition
\(
\mathcal{H}=\mathcal{H}^{(\le n-2)}\oplus\mathcal{H}^{(n-1)} \oplus\mathcal{H}^{(\ge n)}
\),
which corresponds to the block-index partition $I_{n-1} \mid \{n\}\mid I_{n}^{\mathrm{c}}$.
Throughout this subsection, the block index $i$ labels the $(i-1)$-excitation sector.
With respect to this decomposition,
$\bm{W}_n$ is defined by
\begin{equation}
\bm{W}_n
=
\begin{pmatrix}
\bm{1} & \bm{0} & \bm{0} \\
\bm{0} & \bm{1} & \bm{0} \\
\bm{0} & \tilde{\bm{S}}^{(n-1)}_{I_n^{\mathrm c},\,n}\,
\bigl(\tilde{\bm{S}}^{(n-1)}_{n,n}\bigr)^{-1} & \bm{1}
\end{pmatrix}.
\label{eq:Wn_explicit}
\end{equation}
Here the pivot block $\tilde{\bm{S}}^{(n-1)}_{n,n}$ is assumed to be invertible.

The above construction implies the following properties.
For each $m \ge 1$, $\tilde{\bm{S}}^{(m)}$
satisfies:
\begin{enumerate}
\item
For two disconnected clusters $A$ and $B$, the diagonal $(m+1,m+1)$ block of
$\tilde{\bm{S}}^{(m)\,A\cup B}$ is cluster-additive, i.e.,
it is block-diagonal with respect to the decomposition\\
$\mathcal{H}^{(m)}_{A \cup B}=\bigoplus_{p=0}^m (\mathcal{H}_A^{(m-p)}\otimes\mathcal{H}_B^{(p)})$,
so that
\begin{equation}
\tilde{\bm{S}}^{(m)\,A\cup B}_{m+1,m+1}
=
\bigoplus_{p=0}^{m}
\left(
\tilde{\bm{S}}^{(m)\,A}_{m+1-p,m+1-p}
\otimes
\tilde{\bm{S}}^{(m)\,B}_{p+1,p+1}
\right).
\label{eq:cluster_additive_goal}
\end{equation}

\item
All blocks below the diagonal in the $m$th column vanish, i.e.,
\begin{equation}
\tilde{\bm{S}}^{(m)}_{im}=\bm{0},
\qquad i>m .
\label{eq:upper_triangular_goal}
\end{equation}
\end{enumerate}
As a consequence, $\tilde{\bm{S}}^{(m)}$ is block upper-triangular
up to the $m$th column. Furthermore, all diagonal $(n,n)$ blocks with $1 \le n \le m+1$ are block-diagonal
for disconnected clusters.
This block-diagonal structure arises from cancellations
of mixed-sector contributions in the Schur-complement construction.
A proof of these properties by induction on $m$ is given in Appendix~\ref{sec:proof_cluster_add}.

We now show that the diagonal $(m+1,m+1)$ block obtained from this construction
coincides with the HS prescription~\eqref{eq:HS_m_particle}.
In the present formulation, this block arises from a sequence of Schur-complement
reductions that successively eliminate lower excitation-number sectors.
By the nesting (quotient) property of the Schur complement~\cite{Zhang2005},
the result is independent of the order in which lower sectors are eliminated,
provided that all intermediate pivot blocks $\tilde{\bm{S}}^{(n-1)}_{n,n}$ for $1 \le n \le m$
are invertible.
Consequently, the diagonal block obtained after eliminating all lower sectors
is uniquely determined and coincides with Eq.~\eqref{eq:HS_m_particle},
i.e.,
\[
\tilde{\bm{S}}^{(m)}_{m+1,m+1}
=
\bar{\bm{S}}_{m+1,m+1}.
\]

\subsubsection{Minimal-deformation effective Hamiltonians}
\label{sec:min_def_cluster_add}

Once a cluster-additive basis has been fixed,
the block diagonalization of the Hamiltonian within a given excitation sector
is still not unique.
In this subsection, we fix this remaining freedom by imposing
a minimal-deformation criterion on the similarity transformation
acting in that sector.

Specifically, we determine the effective Hamiltonian in the target excitation sector
by selecting, among all transformations that achieve block diagonalization,
the one that minimally distorts the basis states.
This criterion uniquely fixes the effective Hamiltonian
and depends only on the low-energy eigenvalues and eigenvectors within
the chosen sector.

We construct an effective Hamiltonian in the $m$-excitation sector by block-diagonalizing the Hamiltonian
with respect to the excitation-number decomposition in a cluster-additive basis.
To this end, we use an invertible similarity transformation $\bm{W}$ such that the transformed
eigenvector matrix $\tilde{\bm{S}}=\bm{W}^{-1}\bm{S}$ reproduces, in its $(m+1,m+1)$ block,
the HS prescription in Eq.~\eqref{eq:HS_m_particle}.
One explicit realization is the inductive construction
$\bm{W}=\bm{W}_1\bm{W}_2\cdots\bm{W}_m$ introduced above, although the choice of $\bm{W}$ is not unique.

In this basis, the Hamiltonian is $\tilde{\bm{H}}= \bm{W}^{-1} \bm{H} \bm{W}$.
We then apply an additional similarity transformation $\tilde{\bm{T}}$ to
block-diagonalize $\tilde{\bm{H}}$,
\begin{equation}
\tilde{\bm{H}}_{\mathrm{eff}}
=
\tilde{\bm{T}}^{-1}\,\tilde{\bm{H}}\,\tilde{\bm{T}}.
\end{equation}
We focus on the block that decouples the $m$-excitation sector from all remaining sectors.
Let $n$ denote the dimension of the $m$-excitation sector,
so that $\tilde{\bm{T}}_{m+1,m+1}$ is an $n \times n$ matrix.
To minimize the distortion of the cluster-additive basis inside the target sector,
we impose the following criterion.

\begin{criterion}
We choose $\tilde{\bm{T}}$ to block-diagonalize $\tilde{\bm{H}}$ and to satisfy,
within the $m$-excitation sector, the minimal-deformation condition
\begin{equation}
\|\tilde{\bm{T}}_{m+1,m+1} - \bm{1}_n\|_F = \text{minimum}.
\label{eq:min_condition22}
\end{equation}
\end{criterion}

Since $\tilde{\bm{S}}^{(m)}$ diagonalizes $\tilde{\bm H}$,
any similarity transformation that block-diagonalizes $\tilde{\bm H}$
can be written in the form $\tilde{\bm{T}}=\tilde{\bm{S}}^{(m)}\tilde{\bm{F}}$,
where $\tilde{\bm{F}}$ is block-diagonal.
In the following, we restrict to block-diagonal unitary transformations $\tilde{\bm{F}}$,
which preserve Hermiticity and reduce the minimization problem to a well-defined unitary Procrustes problem.
We employ the singular-value decomposition (SVD) of each diagonal block,
\begin{equation}
\tilde{\bm{S}}_{ii}^{(m)}
=
\tilde{\bm{U}}_i \,
\tilde{\boldsymbol{\Sigma}}_i \,
\tilde{\bm{V}}_i^\dagger,
\qquad
i = 1,2,\dots,n_{\max}+1,
\label{eq:SVD_blocks}
\end{equation}
where $\tilde{\bm{U}}_i$ and $\tilde{\bm{V}}_i$ are unitary matrices and
$\tilde{\boldsymbol{\Sigma}}_i$ is diagonal with non-negative entries.

Under the above conditions,
the minimal-deformation criterion uniquely determines the $m$-excitation effective Hamiltonian block
$\tilde{\bm{H}}_{\mathrm{eff},\,m+1,m+1}$.
Importantly, the resulting block $\tilde{\bm{H}}_{\mathrm{eff},\,m+1,m+1}$ depends only
on $\tilde{\bm{S}}_{m+1,m+1}^{(m)}$
and is therefore independent of the particular choice of the similarity transformation $\bm{W}$.

\begin{theorem}\label{thm:2}
Let $\tilde{\bm S}$ be a cluster-additive eigenvector matrix constructed as in Sec.~\ref{subsec:similarity_basis_transformation},
and assume that the block
$\tilde{\bm S}_{m+1,m+1}$ is invertible.
Let $\tilde{\bm{T}}=\tilde{\bm S}\tilde{\bm F}$ be a block-diagonalization transformation satisfying Criterion 2,
with $\tilde{\bm F}$ block-diagonal unitary.

Then the block corresponding to the $m$-particle sector is uniquely determined as
\begin{equation}\label{eq:T22}
\tilde{\bm{T}}_{m+1,m+1} =
\tilde{\bm{U}}_{m+1}\, \tilde{\boldsymbol{\Sigma}}_{m+1}\, \tilde{\bm{U}}_{m+1}^\dagger.
\end{equation}
The corresponding effective-Hamiltonian block is uniquely given by
\begin{equation}
\tilde{\bm{H}}_{\mathrm{eff},\,m+1,m+1}
=
\tilde{\bm{U}}_{m+1}\, \tilde{\bm{V}}_{m+1}^\dagger \,\boldsymbol{\Lambda}_{m+1,m+1}\,
\tilde{\bm{V}}_{m+1}\, \tilde{\bm{U}}_{m+1}^\dagger.
\end{equation}
\end{theorem}

The proof is given in Appendix~\ref{sec:proof}.
This construction yields the same effective-Hamiltonian block as the HS prescription in Ref.~\cite{Hormann2023}, and is uniquely minimal in the sense of Eq.~\eqref{eq:min_condition22}.

\subsection{Numerical selection of target eigenstates}\label{sec:select}
The criteria introduced in Eqs.~\eqref{eq:min_condition} and \eqref{eq:min_condition22}
also provide practical guidelines for selecting eigenstates in numerical calculations.
In particular, they identify the eigenstates that remain most closely aligned with the original basis states.

When the Hamiltonian consists only of $H_0$, the eigenstates belonging to different sectors
are completely separated and can be uniquely distinguished by the quantum numbers used for the sector decomposition.
When a perturbation $V$ that does not conserve these quantum numbers is introduced, such that
$H=H_0+\lambda V$ with a control parameter $\lambda$,
the original eigenstates $|a_j\rangle$ ($j=1,\dots,n$) in the target subspace gradually evolve into
eigenstates $|\phi_j\rangle$ of the full Hamiltonian, which no longer possess definite quantum numbers.
Consequently, each $|\phi_j\rangle$ becomes a linear combination of the basis states $|a_k\rangle$
from both the target sector 
and the remaining sectors. 

As $\lambda$ increases, when two states belonging to different sectors come close in energy,
$V$ can induce strong hybridization between them, leading to avoided level crossings and a subsequent
interchange of eigenstates.
In such cases, the original states continuously evolve into
entirely different ones near the avoided level crossings, where the perturbative expansion breaks
down—a well-known difficulty in perturbative analyses.
The effective Hamiltonian must therefore be constructed using the eigenstates
after the interchange~\cite{Coester2015} that most closely resemble the original target states.

To avoid ambiguity between the two settings, we introduce a unified notation.
Let ${\bm T}_{\mathrm{tar}}$ denote the block of the transformation acting within the target sector:
${\bm T}_{\mathrm{tar}}={\bm T}_{11}$ for the setting of Sec.~\ref{sec:BD_eff_H},
and ${\bm T}_{\mathrm{tar}}=\tilde{\bm T}_{m+1,m+1}$ for the gapped case discussed in Sec.~\ref{sec:gap}.
In both cases, the target eigenstates are selected by minimizing
$\|{\bm T}_{\mathrm{tar}}-{\bm 1}_n\|$ within the candidate subspace.

In the numerical implementation, consider a candidate set of $n$ eigenstates
$\{|\phi_{i}\rangle:i\in \mathcal{S}\}$,
where $\mathcal{S}\subset\{1,\cdots,N\}$ and $|\mathcal{S}|=n$.
For each subset $\mathcal{S}$,
the corresponding block ${\bm T}_{\mathrm{tar}}(\mathcal{S})$ is constructed
(using Eq.~\eqref{eq:T11} for Criterion~1 or Eq.~\eqref{eq:T22} for Criterion~2),
and the quantity $\|{\bm T}_{\mathrm{tar}}(\mathcal{S})-{\bm 1}_n\|$ is evaluated.
The optimal subset $\mathcal{S}^*$ is then determined by
\begin{equation}\label{eq:optimal_eigenS}
  \mathcal{S}^* =
  \arg\min_{\substack{\mathcal{S}\subset\{1,\dots,N\}\\|\mathcal{S}|=n}}
  \|{\bm T}_{\mathrm{tar}}(\mathcal{S})-{\bm 1}_n \|.
\end{equation}
This subset corresponds to the eigenstates that remain most closely aligned with
the original basis states in the target sector.

Section~\ref{sec:2d} demonstrates this selection process in a system exhibiting avoided level crossings
and eigenstate reordering.
Figure~\ref{fig:energy_levels} illustrates the outcome of this state-selection procedure.

\subsection{Numerical linked-cluster expansion (NLCE)}\label{sec:nlc}

For a finite-size cluster $c$, the effective Hamiltonian is obtained by performing block diagonalization
with a suitably chosen set of eigenstates.
The resulting Hamiltonian contains matrix elements representing transitions between local states,
whose magnitudes correspond to cluster-specific physical quantities.
To systematically extrapolate these quantities to the thermodynamic limit,
we employ NLCE~\cite{RigolBS2006,RigolBS2007,Tang2013}, following the procedure outlined in Ref.~\cite{YangS2011}.

We consider a model defined on a lattice with $N$ subunits (e.g., sites, dimers, or
tetrahedra), and aim
to compute the expectation value of an extensive observable ${\cal O}$ per subunit,
$\frac{1}{N} \langle {\cal O} \rangle_c$,
where $\langle {\cal O} \rangle_c$ denotes
the value obtained from a finite cluster $c$ with open boundary conditions.
In the NLCE approach, the $n$th-order estimate of this quantity is given by
\begin{align}\label{eq:NLCE}
 \frac{1}{N} \langle {\cal O} \rangle_{{\rm NLCE}(n)} & = \sum_{c,\,{\rm size}(c)\le n} l_c W_c,
\end{align}
where the summation runs over all inequivalent connected clusters up to size $n$.
Here, the cluster multiplicity $l_c$ denotes
the number of embeddings of cluster $c$ per subunit in the infinite lattice, and
$W_c$ is the cluster weight defined recursively as
\begin{align}\label{eq:NLCweight}
 W_c = \langle {\cal O}\rangle_c - \sum_{s\in c}W_s,
\end{align}
where the summation runs over all connected subclusters $s$ of $c$.

In the present study,
the quantities of interest are the matrix elements of the effective Hamiltonian.
For example, to obtain the NLCE estimate of the nearest-neighbor hopping amplitude, we define
an extensive quantity on each cluster $c$ as
$$
\mathcal{T}_{1,c} = \sum_{i} \sum_{j \in {\rm nn}_i} [{\bm H}_{{\rm eff}, c}]_{ij},
$$
where the inner sum runs over all nearest-neighbors of subunit $i$.
The NLCE estimate of $\frac{\mathcal{T}_1}{N}$ is then obtained using Eqs.~(\ref{eq:NLCE}) and (\ref{eq:NLCweight})
with $\mathcal{T}_1$ in the place
of $\langle {\cal O} \rangle$.
The corresponding hopping amplitude is given by
$$
t_1 = \frac{\mathcal{T}_1}{z N},
$$
where $z$ is the coordination number of the lattice.

\subsection{Asymptotic forms of effective Hamiltonians in gapped systems}\label{sec:asympt}
Before applying the numerical framework to specific models,
we first examine the asymptotic behavior of effective Hamiltonians
in systems with a unique ground state and a finite excitation gap.
This analysis provides physical insight into the rapid convergence of NLCE
and the short-range nature of effective couplings in gapped systems.

We focus on the single-excitation sector on a translationally invariant $d$-dimensional hypercubic lattice.
In this sector, the effective Hamiltonian can be written as
\begin{align}
  H_{\rm eff} & = \sum_{i,j} t_{ij} a_i^\dagger a_j,
\end{align}
where $t_{ij}=t_{ji}^\ast$, and $a_i^\dagger$ ($a_i$) creates (annihilates)
an excitation at site $i$.
The ground state corresponds to the vacuum.
The dispersion relation $\varepsilon ({\bm k})$ is related to $t_{ij}$ through the Fourier transform
 $ \varepsilon ({\bm k}) = \sum_{j} t_{ij} \, \exp \{ i {\bm k}\cdot ({\bm r}_i-{\bm r}_j)\}$,
which is independent of $i$.

To characterize the spatial decay of $t_{ij}$ both near and away from criticality,
we consider the following form of the dispersion relation:
\begin{align}
  \varepsilon ({\bm k}) & = \left[ 2 \left(d-\sum_{\mu=1}^{d}\cos k_\mu \right) +m^{2/z} \right]^{z/2}.
\end{align}
Although this form is not intended to represent any specific model,
it captures the essential low-energy and long-wavelength features
of a generic gapped dispersion in a $d$-dimensional system.
Here, $z$ is the dynamical exponent and $m$ controls the energy gap, satisfying
$\varepsilon ({\bm 0}) = m$. At the critical point $(m=0)$, the low-energy dispersion behaves as
$\varepsilon ({\bm k}) \simeq |{\bm k}|^{z}$.

The Fourier transform of $\varepsilon ({\bm k})$ determines the asymptotic behavior of $t_{ij}$.
For $z\neq 2$, we obtain
\begin{equation}
t_{ij} \propto
\begin{cases}
\exp\!\left(-m\,|\bm{r}_i-\bm{r}_j|\right),
& |\bm{r}_i-\bm{r}_j|\gg m^{-1},\\[3pt]
|\bm{r}_i-\bm{r}_j|^{-d-z},
& |\bm{r}_i-\bm{r}_j|\ll m^{-1},
\end{cases}
\end{equation}
whereas for $z=2$ the couplings are strictly short-ranged,
\begin{equation}
t_{ij} \propto
\sum_{\mu=1}^{d}
\left(
\delta_{\bm{r}_i-\bm{r}_j,\bm{e}_\mu}
+
\delta_{\bm{r}_i-\bm{r}_j,-\bm{e}_\mu}
\right),
\end{equation}
with $\bm{e}_\mu$ the unit lattice vectors.

These results illustrate that, for dispersions of the above generic gapped form, the effective couplings decay exponentially with distance.
Consequently, NLCE converges rapidly even when truncated to relatively small clusters,
enabling an accurate and efficient construction of effective Hamiltonians.

\section{Application I: one-dimensional transverse-field Ising model}
\label{sec:1d}

As a first application, we consider magnon excitations in the one-dimensional transverse-field Ising model.
The Hamiltonian is defined as
\begin{align}\label{eq:model}
H_\text{1d} = - \sum_{i=1}^{n} \sigma_i^z -J\sum_{i=1}^{n-1} \sigma_i^x \sigma_{i+1}^x,
\end{align}
where $\sigma_i^\alpha$ denotes the Pauli matrix at site $i$.
The first term defines the unperturbed Hamiltonian $H_0$, which aligns all spins along the field direction,
while the second term acts as the perturbation $V$ describing the transverse spin-exchange interaction
between neighboring sites.
The system consists of an $n$-spin chain (cluster) $c$ with open boundary conditions.

We consider the parameter range $0 \le J \le 1$.
For $0 \le J < 1$, the ground state is unique and polarized along the field direction.
This exactly solvable model~\cite{Pfeuty1970} has previously served as a benchmark for the continuous unitary
transformation approach~\cite{YangS2011}.
Our results are directly compared with those earlier results, as discussed in
Appendix~\ref{sec:comp}.

We derive the effective Hamiltonian in the one-magnon sector.
In the unperturbed Hamiltonian, the target subspace is defined by fixing
the spin-flip quantum number $S^z = n/2 - 1$,
which uniquely determines an $n$-dimensional subspace
for an $n$-site cluster.
The Hilbert space of the unperturbed system is partitioned into three sectors based on the number of spin flips (equivalently, the total $S^z$):
(i) the fully polarized state with $S^z=n/2$,
(ii) an $n$-dimensional subspace of single spin-flip states with $S^z=n/2-1$,
and (iii) the remaining higher-excitation subspace.
The Hamiltonian is block-diagonalized with respect to this decomposition.
In practical calculations, we diagonalize the full Hamiltonian $H$
on each finite cluster and select, from the low-energy spectrum,
the $n$ eigenstates that minimize the deviation measure
in Eq.~\eqref{eq:optimal_eigenS}
with respect to the unperturbed one-magnon basis.
The effective Hamiltonian is then constructed from the block of the
selected $n$ eigenstates corresponding to the one-magnon basis.

Because the perturbation $V$ changes the number of spin flips only by
$0$ or $\pm 2$, the one-magnon eigenstates do not couple to the
unperturbed ground state.
This situation corresponds to the case discussed in Sec.~\ref{sec:BD_eff_H},
where no additional basis transformation is required and
the CSM block-diagonalization procedure can be applied directly
within the fixed one-magnon subspace.

In the first sector,
the diagonal element of the block-diagonalized Hamiltonian
reduces to the ground-state energy,
\begin{align}\label{eq:H0}
  H_{\rm eff}^{(0)} & = E_0(n).
\end{align}
In the second single-spin-flip sector, the effective Hamiltonian takes the form
\begin{align}\label{eq:H1}
  H_{\rm eff}^{(1)} & =  E_0(n) + \sum_{i, j\in c} t_{ij}(n) a_i^\dagger a_j,
\end{align}
where $t_{ij}(n)=t_{ji}(n) \in \mathbb{R}$,
and $a_i$ ($a_i^\dagger$) denotes the annihilation (creation) operator of a single spin flip at site $i$.
The ground–state energy term in Eq.~(\ref{eq:H1}) is retained explicitly, as required
by the irreducible construction in Eq.~(\ref{eq:subtract}),
to ensure that the excitation-energy contributions remain cluster additive~\cite{Gelfand1996}.

We evaluate the ground-state energy $E_0(n)$ and the distance-$r$ summed hopping amplitudes
$\sum_{i=1}^{n-r} t_{i,i+r}(n)$ for $r=0,\dots,n-1$ on each cluster.
These quantities are then used as input for NLCE to obtain the corresponding
local couplings $t_{i,i+r}^{\rm NLCE}$ in the thermodynamic limit, which are translationally invariant.
The expansion is performed using clusters up to $n=16$.

\begin{figure}
  \centering
  \includegraphics[width=10cm]{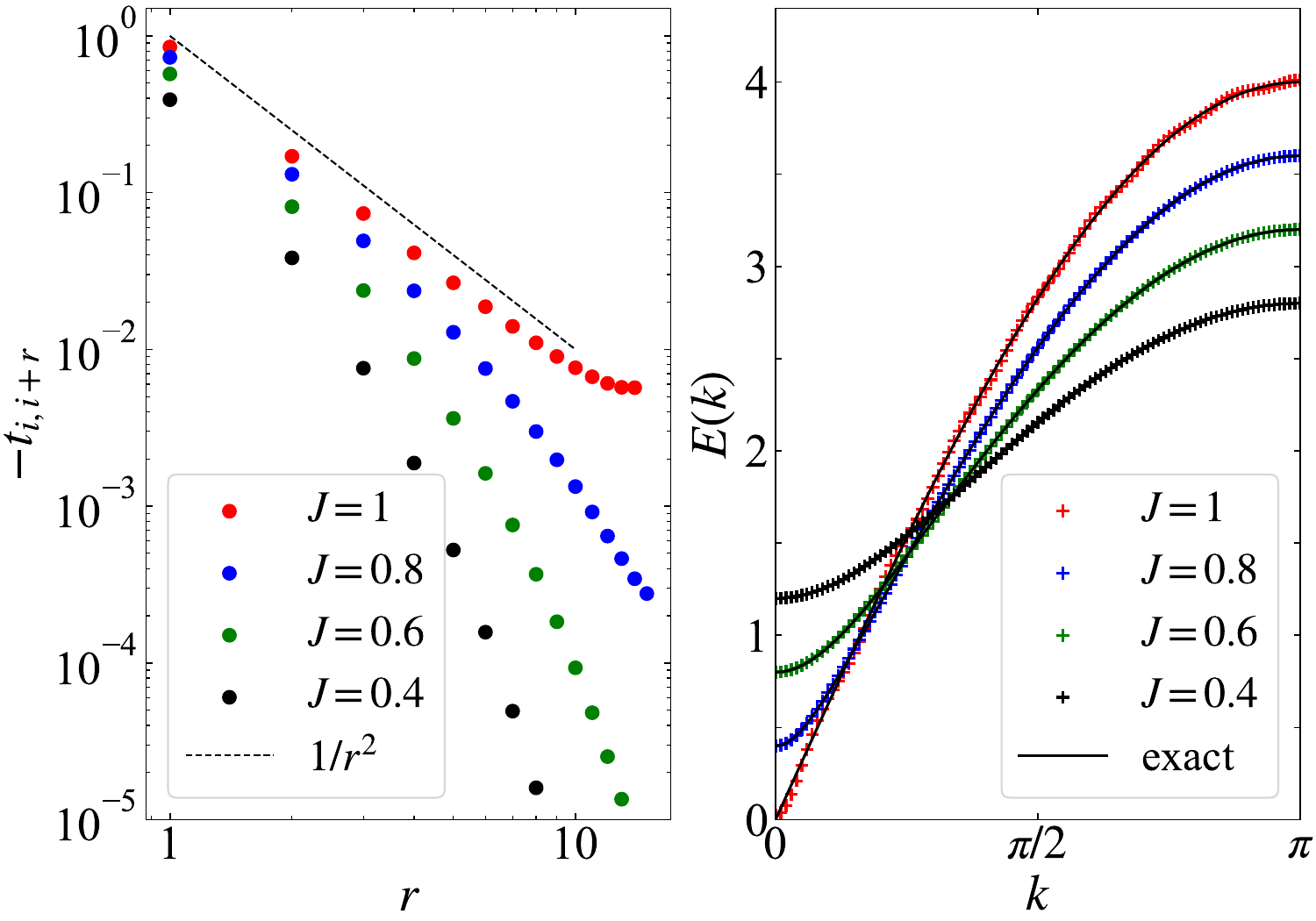}
  \caption{Numerical results obtained from the block diagonalization and numerical linked-cluster expansion.
  (Left) Distance dependence ($r$-dependence) of the magnon hopping amplitudes $t_{i,i+r}^{\rm NLCE}$
  (independent of $i$).
  (Right) One-magnon excitation spectra compared with the exact results.}\label{fig:hopping}
\end{figure}

Figure~\ref{fig:hopping} shows the estimated hopping amplitudes $t_{i,i+r}^{\rm NLCE}$ as a function of distance $r$,
and the corresponding one-magnon excitation spectrum.
For $J<1$, the hopping amplitudes $t_{i,i+r}^{\rm NLCE}$ decay exponentially with $r$, consistent
with a finite excitation gap.
As $J$ approaches 1, the decay becomes slower, and at $J=1$ it follows a $1/r^2$ power-law behavior,
indicating a gapless excitation spectrum with dynamical exponent $z=1$.

Despite the limited cluster size considered (up to $n=16$),
the resulting excitation spectra show excellent agreement with the exact results,
demonstrating the quantitative accuracy of the present approach and its consistency with previous studies~\cite{YangS2011}.

\section{Application II: two-dimensional Shastry--Sutherland model with Dzyaloshinskii--Moriya couplings}
\label{sec:2d}
As a second application, we study triplon excitations
in the two-dimensional Shastry--Sutherland model~\cite{shastry1981}
with Dzyaloshinskii--Moriya (DM) interactions,
motivated by SrCu$_2$(BO$_3$)$_2$~\cite{Kageyama1999}.

\subsection{Model}
\subsubsection{Microscopic spin model}
\begin{figure}
  \centering
  \includegraphics[width=9.5cm]{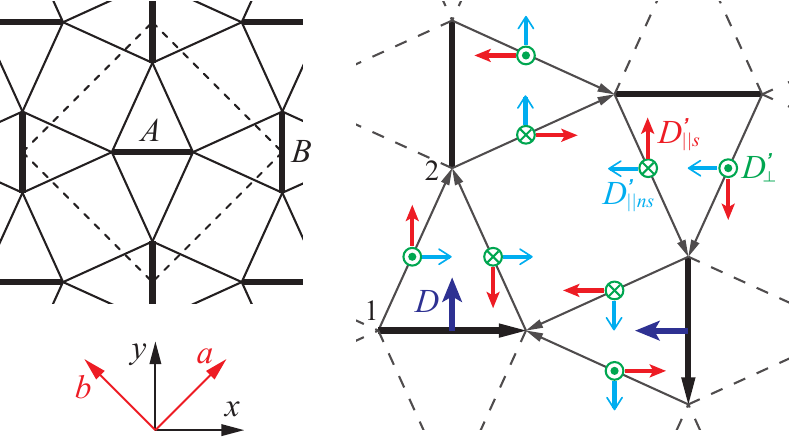}
  \caption{(Left) Unit cell of the Shastry--Sutherland lattice.
  (Right) Symmetry-allowed configurations of DM vectors in SrCu$_2$(BO$_3$)$_2$, consistent with $S_4$ and
  $C_{2V}$ crystal symmetries.
  The site indices of ${\bm D}_{ij}$ and ${\bm D}^\prime_{ij}$ are assigned
  following the bond direction $i\rightarrow j$, indicated by the arrows.
  }\label{fig:SCBO}
\end{figure}

The model is defined on a two-dimensional lattice consisting of orthogonal $A$- and $B$-type dimers,
as illustrated in Fig.~\ref{fig:SCBO}.
The total Hamiltonian is expressed as $H_{\rm 2d} = H_{\rm D} + H_{\rm ID}$,
where $H_{\rm D}$ describes the intradimer interactions:
\begin{align}\label{eq:Hamiltonian_intra}
  H_{\rm D} & = \sum_{\langle i,j \rangle_1} (J\,{\bm S}_i \cdot {\bm S}_j
  + {\bm D}_{ij}\cdot {\bm S}_i \times {\bm S}_j )
  -h\sum_{j} S_j^z ,
\end{align}
and $H_{\rm ID}$ represents the interdimer couplings:
\begin{align}\label{eq:Hamiltonian_inter}
  H_{\rm ID} & = \sum_{\langle i,j \rangle_2} (J'\, {\bm S}_i \cdot {\bm S}_j
  + {\bm D}'_{ij} \cdot {\bm S}_i \times {\bm S}_j) .
\end{align}
Here, $\langle i,j \rangle_n$ denotes $n$th-neighbor spin pairs.

The DM vectors ${\bm D}_{ij}$ and ${\bm D}'_{ij}$ are constrained by the $I\bar{4}2m$ space-group
symmetry~\cite{Sparta2001} of SrCu$_2$(BO$_3$)$_2$,
which contains an $S_4$ symmetry centered on the square plaquettes of four dimers and
a $C_{2V}$ symmetry centered on each individual dimer.
The corresponding symmetry-allowed configurations of DM vectors are shown in Fig.~\ref{fig:SCBO}.
We adopt the parameter set ${\bm D}_{ij}/J=(0,0.048,$ 0) for $A$-type dimers and
${\bm D}_{ij}^\prime/J\equiv(D_{\parallel{\rm ns}}^\prime,D'_{\parallel{\rm s}},D'_{\perp})=(0.005,0.008,0.014)$
for the (1,2) bond shown in Fig.~\ref{fig:SCBO}, following
\textit{ab initio} calculations~\cite{Marurenko2008}.

\subsubsection{Effective triplon model}\label{sec:eff_tripl_H}

To derive the effective Hamiltonian, we choose the unperturbed part as
\begin{align}
  H_0 = J \sum_{\langle i,j \rangle_1} {\bm S}_i \cdot {\bm S}_j - h \sum_j S_j^z,
\end{align}
and treat the remaining interactions as the perturbation $V$, so that $H_\mathrm{2d} = H_0 + V$.
In the low-field regime, $H_0$ has the dimer-singlet product ground state, where all dimers form singlets,
and its excited states are triplon configurations.
Since $H_0$ conserves the triplon number, the Hilbert space decomposes into sectors
with fixed triplon number.
We block-diagonalize $H_\mathrm{2d}$ in the one-triplon sector
and derive the corresponding effective Hamiltonian.

Each dimer is characterized by its center position ${\bm r}$, with its two spins labeled as sites 1 and 2.
Triplon operators, which create or annihilate triplon excitations, are written as
$t^{\mu\dagger}_{\bm r} = i S_{{\bm r},1}^\mu - i S_{{\bm r},2}^\mu$,
$t^\mu_{\bm r} = -i S_{{\bm r},1}^\mu + i S_{{\bm r},2}^\mu$,
for $\mu=x,y,z$, where ${\bm r}$ runs over the lattice $\Lambda$ of all dimer centers.

The effective one-triplon Hamiltonian has the form
\begin{align}\label{eq:eff_H_SS}
   H_{\rm eff}^{(1)} & = E_0+\sum_{\alpha=A,B}\sum_{{\bm r}\in \Lambda_\alpha}
   \left( \sum_{{\bm \delta}}
  {{\bm t}^{\dagger}_{{\bm r}+{\bm \delta}}} {\bm M}_{\alpha\alpha,{\bm \delta}} {\bm t}_{{\bm r}}
  + \sum_{{\bm \delta}'}
  {{\bm t}^{\dagger}_{{\bm r}+{\bm \delta}'}} {\bm M}_{\bar{\alpha}\alpha,{\bm \delta}'} {\bm t}_{{\bm
  r}}\right),
\end{align}
where $E_0$ denotes the ground-state energy, ${\bm t}_{\bm r}= (t^x_{{\bm r}}, t^y_{{\bm r}},t^z_{{\bm r}} )^T$, and
${\bm M}_{\alpha \beta,{\bm \delta}}$ are $3\times 3$ hopping matrices describing transitions of triplons
from $\beta$-type to $\alpha$-type dimers. We define $\bar{A}=B$ and $\bar{B}=A$.
The sublattices $\Lambda_A$ and $\Lambda_B$ refer to the sets of $A$- and $B$-type dimer centers, respectively.
The vectors ${\bm \delta}$ (${\bm \delta}'$) denote relative positions within (between)
sublattices.
Hermiticity imposes the condition
${\bm M}^{\dagger}_{\alpha \beta,{\bm \delta}}={\bm M}_{\beta \alpha,-{\bm \delta}}$.

Lattice symmetries impose additional constraints on the matrices
${\bm M}_{\alpha \beta,{\bm \delta}}$
(see Appendix~\ref{sec:sym_const}).
Enforcing these symmetries at the level of the effective Hamiltonian
substantially reduces the number of independent hopping parameters.
The remaining parameters behave as extensive quantities and therefore
provide suitable input for NLCE extrapolation.

\begin{figure}[t]
  \centering
  \includegraphics[width=8.5cm]{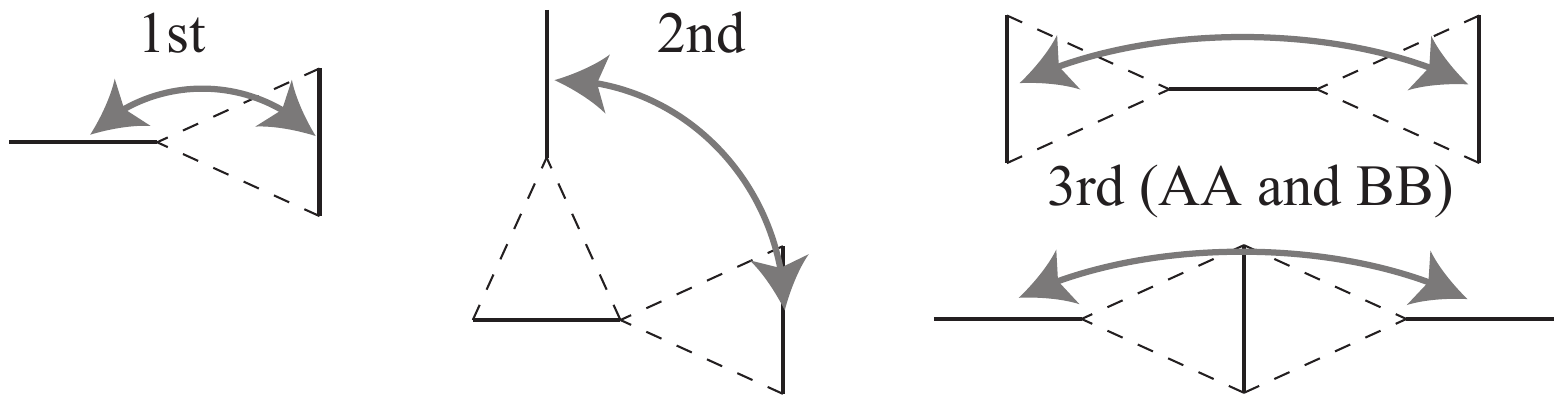}
  \caption{
Schematic illustration of the hopping paths connecting first, second, and third-neighbor dimers and the corresponding minimal
connected clusters used in the NLCE analysis.}\label{fig:path}
\end{figure}

For instance, the on-site potential ${\bm M}_{AA,(0,0)}$ and the nearest-neighbor hopping matrix ${\bm M}_{BA,(1,0)}$
take the forms
\begin{align}\label{eq:M00M01}
M_{AA,(0,0)} =
                     \begin{bmatrix}
                       R_{01} & I_{01} & 0 \\
                      -I_{01} & R_{02} & 0 \\
                       0 & 0 & R_{03} \\
                     \end{bmatrix},\quad\quad
{\bm M}_{BA,(1,0)}=
            \begin{bmatrix}
              I_{11} & R_{11} & I_{12} \\
              R_{12} & I_{13} & R_{13} \\
              I_{14} & R_{14} & I_{15} \\
            \end{bmatrix},
\end{align}
where $R_i$ and $I_i$ are real and imaginary coefficients, respectively.
In the present analysis, triplon hopping processes up to second and third neighbors are included.
The corresponding hopping paths, together with the minimal connected clusters used in the NLCE analysis,
are illustrated in Fig.~\ref{fig:path}.

\subsection{Numerical results}

\subsubsection{Basis transformation and block diagonalization}

The $J'$ interaction alone does not modify the singlet ground state for
$J'/J \lesssim0.675$~\cite{shastry1981,CorbozM2013}.
However, once the DM interactions ${\bm D}$ and ${\bm D}'$ are included, the ground state acquires
a finite admixture beyond the pure singlet configuration.
The one-triplon eigenstates also acquire admixtures of the unperturbed singlet ground
state~\cite{Cepas2001,Room2004,Kodama2005,Romhanyi2011},
and this situation corresponds to the case in Sec.~\ref{sec:min_def_cluster_add},
where a basis transformation is required to remove admixtures of the ground state prior to block diagonalization.
Therefore, in our actual calculations we first construct the transformed eigenvector block
$\tilde{\bm S}_{22}$ according to Eq.~\eqref{eq:HS_condition_22}.
Using this block as input, we then build the effective Hamiltonian from Theorem~\ref{thm:2}.

\subsubsection{Eigenstate selection}

Because the perturbation $V$ does not conserve the triplon number,
eigenstates belonging to different triplon-number sectors can exhibit
avoided level crossings and exchange their eigenstate character
when they approach each other in energy.
To construct the effective Hamiltonian, we therefore select the eigenstates in the target subspace
according to the minimization condition \eqref{eq:min_condition22} (see Sec.~\ref{sec:select}),
ensuring that they remain closest to the original one-triplon basis.

To illustrate avoided crossings and the eigenstate selection procedure,
we introduce a parameter $\lambda$ for the interdimer term.
The deformed Hamiltonian is defined as
\begin{align}\label{eq:lambda}
  H_{\rm 2d}(\lambda) & = H_{\rm D} +\lambda H_{\rm ID}.
\end{align}
The original Hamiltonian for SrCu$_2$(BO$_3$)$_2$ corresponds to $\lambda=1$.

\begin{figure}[t]
  \centering
    \begin{subfigure}[t]{0.75\linewidth}
    \centering
    \caption{}\label{fig:energy_levels:a}
    \includegraphics[width=\linewidth]{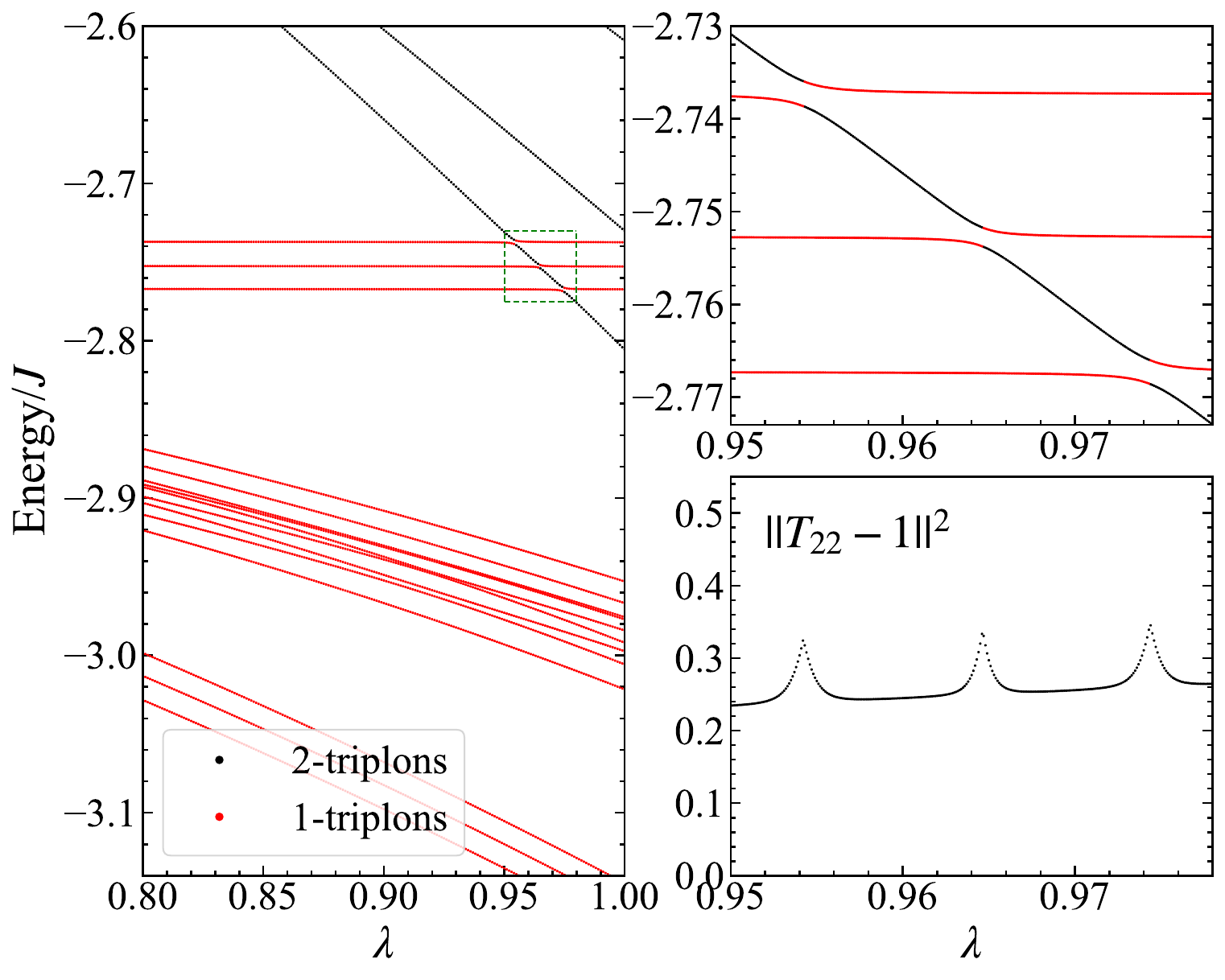}
  \end{subfigure}\hfill
  \begin{subfigure}[t]{0.22\linewidth}
    \centering
    \caption{}\label{fig:energy_levels:b}
    \includegraphics[width=\linewidth]{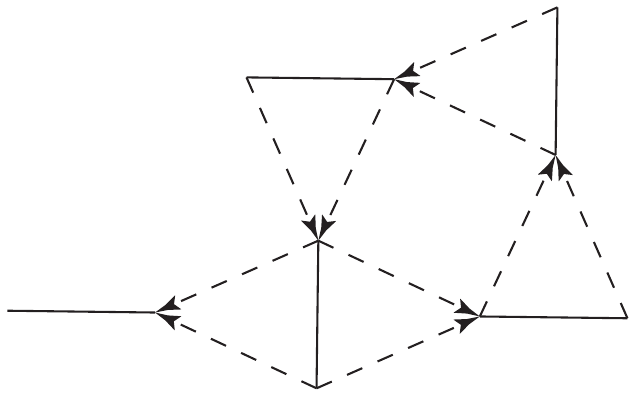}
  \end{subfigure}
  \caption{(a) Left: $\lambda$-dependence of the energy spectrum for the 10-spin cluster shown in (b).
  Red dots correspond to one-triplon excitations, and black dots represent two-triplon states.
Right (top): Enlarged view of the green dashed region in the left panel.
Following level repulsion, the states interchange, with red lines indicating the eigenstates selected
according to the criterion in Eq.~(\ref{eq:min_condition22}).
Right (bottom): Plot of $||T_{22} - 1||^2$ for the selected eigenstate set.
(b) The 10-spin cluster used in (a).}\label{fig:energy_levels}
\end{figure}

Figure~\ref{fig:energy_levels} illustrates a typical example of the eigenstate selection procedure
as $\lambda$ varies,
where level repulsion and eigenstate interchange occur between one-triplon and two-triplon states.
At each avoided crossing, the character of the two eigenstates is exchanged.
The selection criterion identifies
the eigenstates that retain the character of the original one-triplon basis.

Because the index of the selected eigenstate may jump at such crossings,
this selection process can produce cusp-like anomalies in physical quantities.
The CUT framework provides a method for suppressing such
anomalies~\cite{Coester2015}, whereas no analogous technique is currently available
for other approaches~\cite{Hormann2023}.
Each anomaly typically arises from the exchange of a single pair of eigenstates,
and hence its amplitude does not scale with the system size $n$.

In practice, NLCE substantially mitigates these anomalies and yields smoother physical quantities.
In our numerical implementation, the anomalies are suppressed by incorporating many subclusters
with small size differences generated systematically using
the bond-dilution procedure described below.

\subsubsection{NLCE results}

\begin{figure}[t]
\centering
\begin{minipage}[t]{0.05\linewidth}
\vspace{0pt} (a)
\end{minipage}%
\begin{minipage}[t]{0.75\linewidth}
\vspace{0pt}\includegraphics[width=10.cm]{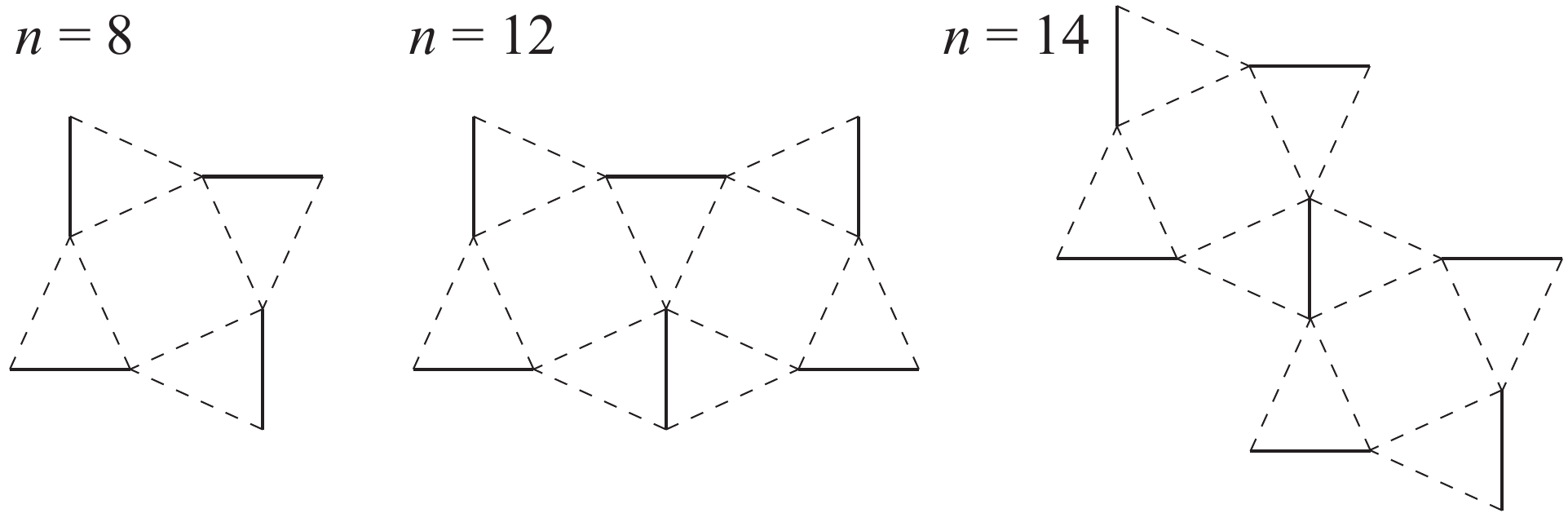}
\end{minipage}

\vspace{6pt}

\begin{minipage}[t]{0.05\linewidth}
\vspace{0pt} (b)
\end{minipage}%
\begin{minipage}[t]{0.75\linewidth}
\vspace{0pt}\includegraphics[width=10.3cm]{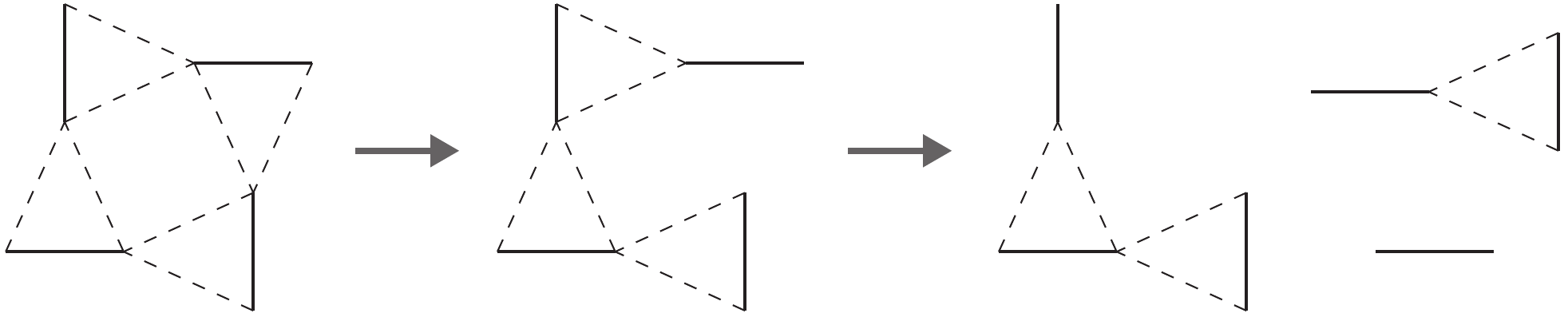}
\end{minipage}

\caption{
(a) Series of clusters constructed from basic unit blocks ($n=8$, four orthogonal dimers).
Larger clusters are obtained by connecting these blocks through edge- or corner-sharing.
(b) Example for $n=8$ illustrating the iterative generation of subclusters,
up to two successive bond-pair removals.
Among the generated clusters, each linked (connected) cluster is taken as a subcluster in the NLCE calculation.
}
\label{fig:clusters}
\end{figure}

Using the eigenstates selected by the above criterion,
we compute the matrix elements of the effective Hamiltonian
and evaluate them using NLCE.
For each hopping term, the expansion begin with the smallest connected cluster containing
the relevant sites (see Fig.~\ref{fig:path}):
one dimer for on-site terms,
two dimers for nearest-neighbor hoppings, and
three dimers connected through interdimer couplings for second- and third-neighbor hoppings.
Higher-order corrections from larger clusters are then incorporated systematically using
the cluster weights defined in Eq.~(\ref{eq:NLCE}),
up to a chosen truncation order.

To construct such larger clusters, we first define a basic unit block consisting of four dimers arranged
in a square
and combine these blocks through edge or corner sharing (Fig.~\ref{fig:clusters}(a)).
Subclusters used in NLCE [Eqs.~(\ref{eq:NLCE}) and (\ref{eq:NLCweight})] are
generated systematically by a bond-dilution procedure that removes selected pairs of interdimer bonds successively
(see Fig.~\ref{fig:clusters}(b)).
In the present calculations, clusters of up to two basic unit blocks (seven dimers, 14 spins) were included.


\begin{figure}[tb]
  \centering
  \begin{subfigure}[t]{0.49\linewidth}
    \centering
    \includegraphics[width=\linewidth]{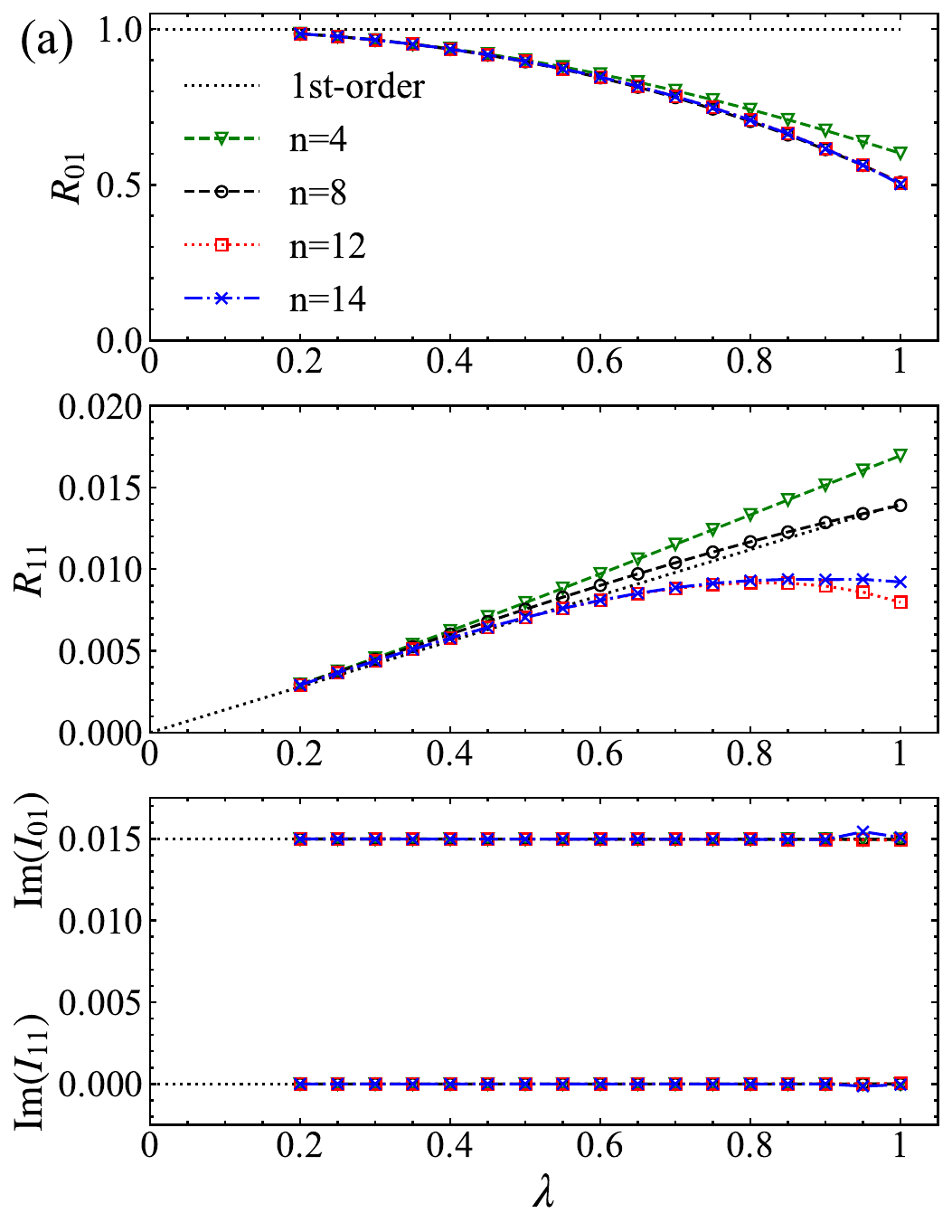}
  \end{subfigure}\hfill
  \begin{subfigure}[t]{0.49\linewidth}
    \centering
    \includegraphics[width=\linewidth]{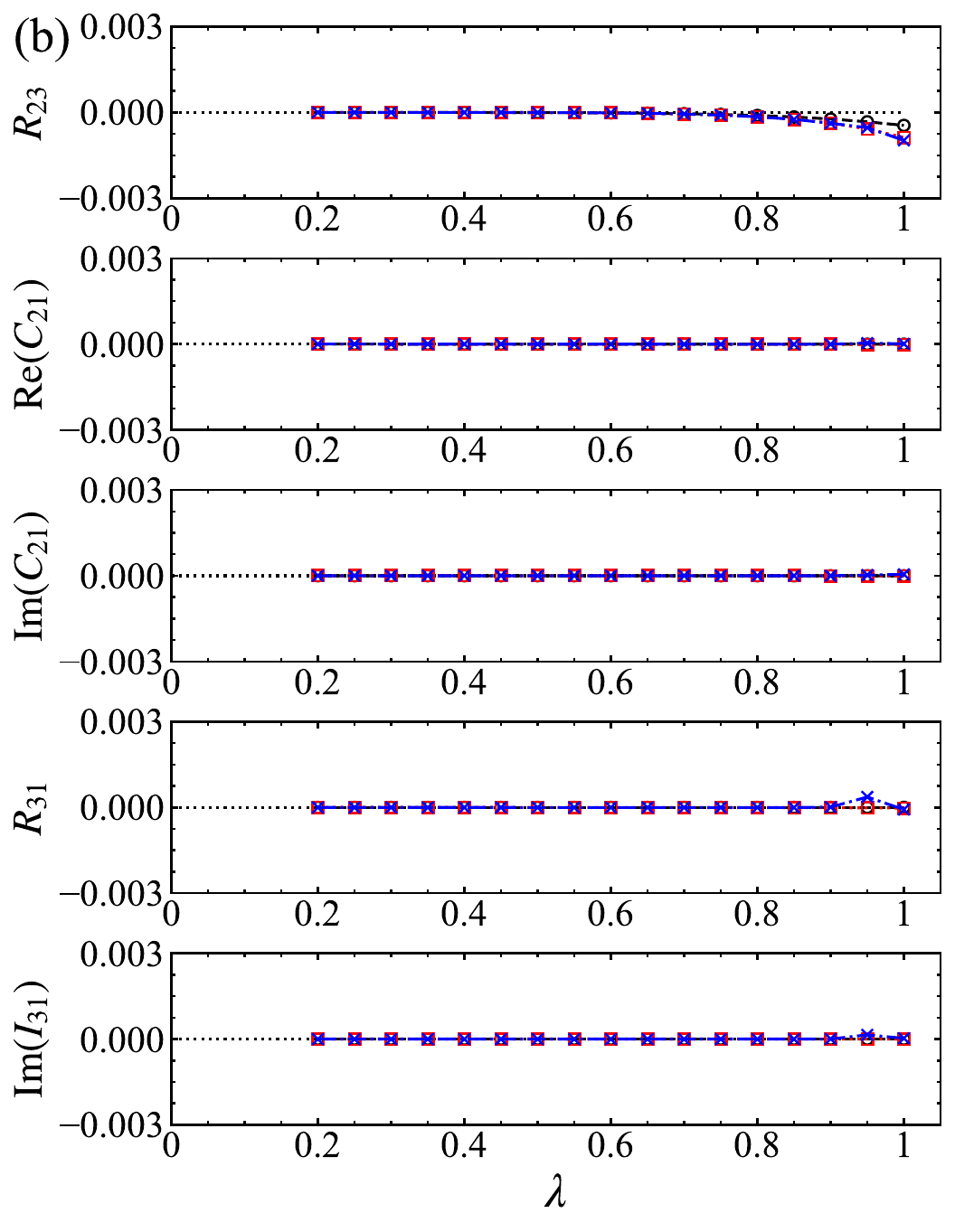}
  \end{subfigure}
  \caption{$\lambda$-dependence of selected matrix elements for $J'/J=0.6$ and $h/J=0.015$.
  (a) On-site potential ($R_{01}$ and $I_{01}$) and nearest-neighbor hopping ($R_{11}$ and $I_{11}$).
  (b) Second-neighbor hopping [$R_{23}$, ${\rm Re}(C_{21})$, and ${\rm Im}(C_{21})$]
  and third-neighbor hopping ($R_{31}$ and $I_{31}$).
  Dashed lines represent the first-order perturbative expansion results~\cite{Romhanyi2015}.
  Here, $n$ denotes the number of spins in the largest clusters used for the numerical linked-cluster expansion.}\label{fig:matrix1}
\end{figure}

To evaluate deviations from perturbative results, we analyze the $\lambda$-dependence of the matrix elements.
Figure~\ref{fig:matrix1}(a) shows the dependence on $\lambda$ and cluster size for selected matrix elements
of the on-site potential and nearest-neighbor hopping, whereas
Fig.~\ref{fig:matrix1}(b) presents the corresponding results for the second-
and third-neighbor hopping.
At $\lambda=1$, the real parts of the on-site potential and nearest-neighbor hopping terms
exhibit significant renormalization,
while their imaginary parts remain nearly unchanged.
Matrix elements associated with second- and third-neighbor hopping are generally
negligible, except for weak enhancements in the diagonal elements of second-neighbor hopping matrices
near $\lambda=1$.
As discussed below, the enhancement of the second–neighbor hopping amplitudes
give rise to the finite dispersion observed in the triplon bands.

\subsubsection{Triplon bands}

\begin{figure}
  \centering
  \includegraphics[width=11.5cm]{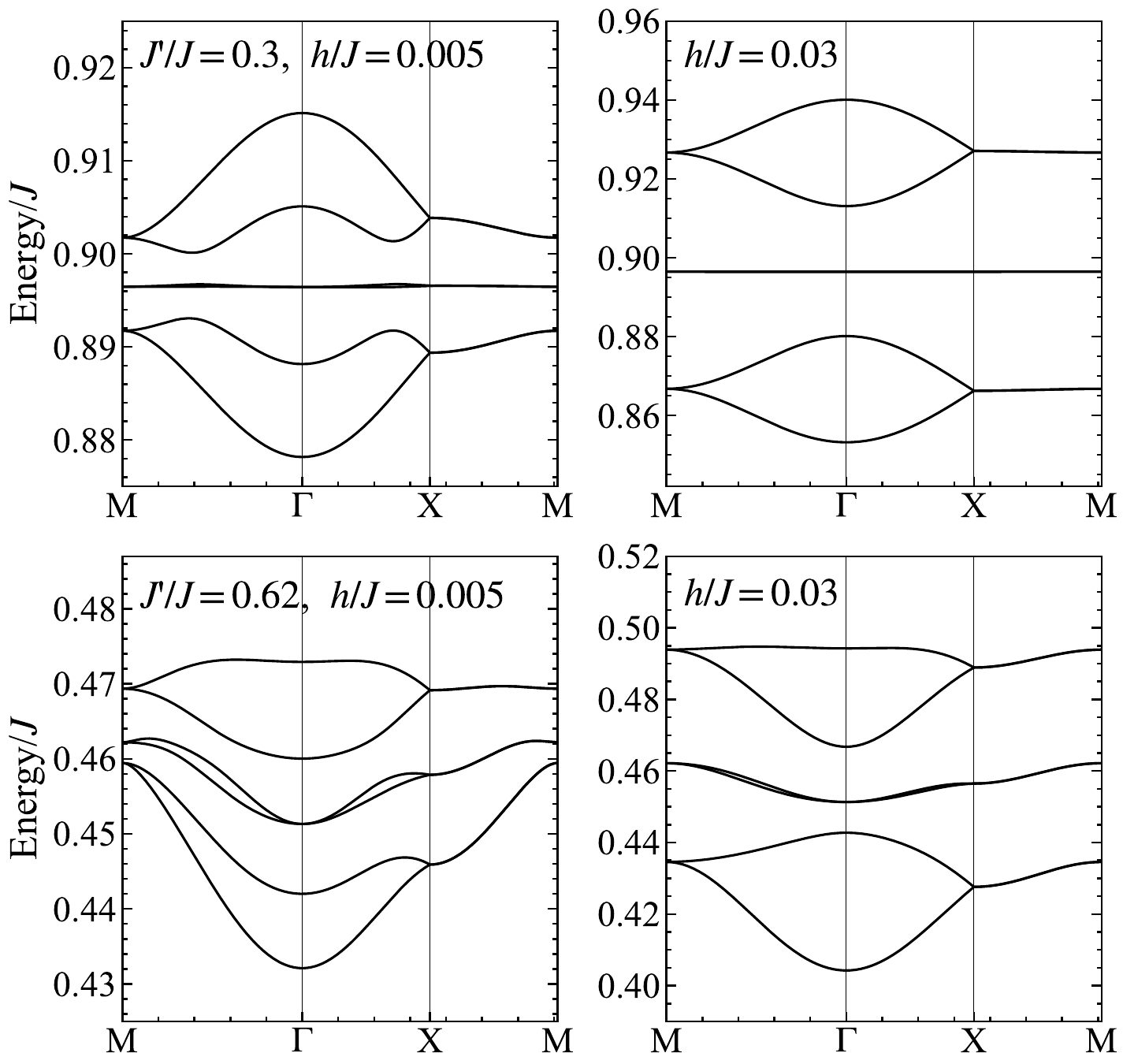}
  \caption{Excitation-energy spectra obtained from the triplon effective Hamiltonians
  for $J' =0.3 J$ (top panels) and $J'=0.62J$ (bottom panels).
  At $h=0.005J$, the triplon bands are topological in both cases, characterized by Chern numbers
  $(C_1,C_2,C_3)=(2,0,-2)$ from the lowest to the highest band.
  At $h=0.03J$, all bands are topologically trivial.
  High-symmetry points in the Brillouin zone are defined as $\Gamma=(k_x,k_y)=(0,0)$, ${\mathrm X}=(\pi/2,\pi/2)$,
  and ${\mathrm M}=(\pi,0)$, where momentum is given in units of the inverse nearest-neighbor
  dimer distance.}
  \label{fig:band_NLC}
\end{figure}

Figure~\ref{fig:band_NLC} shows the excitation-energy spectra
of the triplon bands for $J'/J=0.3$ and $J'/J=0.62$ under different magnetic fields.
Because the unit cell contains two dimers, the system hosts six triplon bands.

Time-reversal symmetry in combination with nonsymmorphic lattice symmetries enforce
symmetry-protected band degeneracies along the Brillouin-zone boundary,
where pairs of bands remain connected.
This behavior is clearly visible in the X-M segment in the figure,
effectively resulting in three connected band pairs.
The symmetry origin of this degeneracy is discussed in Appendix~\ref{sec:Kramers}.
Its reproduction by the effective Hamiltonian provides
a consistency check of the construction.

For $J^\prime=0.3J$, the triplon spectrum exhibits a nearly flat central band,
with the upper and lower bands symmetrically arranged around it,
consistent with perturbative calculations~\cite{Romhanyi2015,Malki2017}.
For $J^\prime=0.62J$, the central band develops a significant dispersion
and the symmetric structure is lost.
This behavior agrees with inelastic neutron-scattering observations~\cite{McClarty2017}.
Whereas previous analyses introduced phenomenological second-neighbor hoppings
to reproduce these features,
the present non-perturbative construction captures them
directly from the microscopic model.

To characterize the band topology, we compute the Berry curvature
\begin{equation*}
\Omega_n ({\bm k}) = \partial_x A_n^y ({\bm k}) - \partial_y A_n^x ({\bm k}),\qquad
A_n^\mu ({\bm k}) =i \langle n({\bm k})|\partial_\mu |n({\bm k}) \rangle,
\end{equation*}
numerically~\cite{Fukui2005},
where $\partial_\mu \equiv \partial/\partial k_\mu$, and $|n({\bm k}) \rangle$ denotes
the normalized eigenstate of the $n$th Bloch band.
The Chern number of the $n$th band is given by
\begin{align}\label{eq:Ch}
  C_n & = -\frac{1}{2\pi} \int d^2k \, \Omega_n ({\bm k}).
\end{align}

Topological triplon bands with nonzero Chern numbers appear in the small-field region \\
$h/J \lesssim 0.02$ for both $J'/J=0.3$ and $0.62$.
For $J'/J=0.3$, we obtain
$(C_1,C_2,C_3)=(2,0,-2)$
in this regime and trivial bands at larger fields,
in agreement with perturbative studies~\cite{Romhanyi2015,Malki2017}.
For $J'/J = 0.62$, the low-field region exhibiting $(2,0,-2)$ becomes narrower,
and additional intermediate topological bands, such as 
$(1,1,-2)$ at $h/J=0.01$ and $(0,1,-1)$ at $h/J=0.015$,  emerge.
These features are not captured in perturbative treatments.

Experimentally, two–triplon excitations and bound states have been observed
in ESR~\cite{Nojiri2003}, inelastic neutron scattering~\cite{McClarty2017} and Raman spectroscopy~\cite{Wulferding2021} experiments on SrCu$_2$(BO$_3$)$_2$.
The present calculations are restricted to the one–triplon sector
and therefore do not address these multi–particle channels.
When such two-triplon bound states soften and close the excitation gap,
they have been suggested to give rise to a spin–nematic phase
in high magnetic fields~\cite{Momoi2000,WangB,Imajo,fogh2023}.
Extending the present non–perturbative framework
to treat two–triplon bound states on an equal footing
with single–triplon excitations
remains an important direction for future work.

\section{Summary and methodological outlook}\label{sec:summary}

We develop a non-perturbative framework for constructing effective Hamiltonians
in quantum many-body systems. The method combines numerical block diagonalization on finite clusters
with extrapolation to the thermodynamic limit using the numerical linked-cluster
expansion (NLCE).
The construction is guided by a minimal-deformation principle: among the many transformations
that decouple a chosen target subspace, we select the transformation that modifies the target
low-energy basis as little as possible.

For quasi-degenerate low-energy manifolds, cluster additivity is automatically satisfied,
and the minimal-deformation criterion directly determines a unique effective Hamiltonian.
For gapped systems with a unique ground state, cluster additivity must first be restored
by constructing a cluster-additive basis.
By reproducing the state-level deformation introduced by H\"ormann and Schmidt (HS)
and subsequently imposing a minimal-deformation criterion within this basis,
the remaining freedom of the block-diagonalizing transformation is fixed in a well-defined manner.
The resulting effective Hamiltonian is equivalent to that obtained by the HS construction,
while the present formulation makes explicit the minimality principle underlying it.
The same criterion also provides a practical eigenstate-selection rule
in the target subspace [cf.~Eq.~(\ref{eq:optimal_eigenS})],
which allows one to track physically relevant eigenstates as parameters vary.
Although the optimal assignment may change near avoided level crossings,
the resulting effective Hamiltonian remains well defined.

We demonstrate the framework using two spin models.
For the one-dimensional transverse-field Ising model,
the one-magnon dispersion is reproduced in agreement with the exact solution.
For the two-dimensional Shastry--Sutherland model with Dzyaloshinskii--Moriya interactions,
the method captures the renormalization of triplon hopping amplitudes
and yields triplon band structures consistent with experimental observations in SrCu$_2$(BO$_3$)$_2$.

A notable feature of the approach is that it relies solely on low-energy eigenstates
and does not require high-energy spectral information.
The minimal-deformation criterion also helps maintain numerical stability
in parameter regimes where perturbative expansions become unreliable,
for instance due to strong level mixing,
avoided level crossings, or eigenstate interchange.
When combined with NLCE, correlations from increasingly large clusters
are incorporated systematically,
enabling the thermodynamic-limit effective Hamiltonian
to encode long-range effective couplings without phenomenological input.

The present work focuses on systems with a unique ground state
and well-defined gapped excitations.
Previous CUT-based NLCE studies~\cite{YangACLS2012,Ixert2014,Ixert2016}
suggest that related constructions can remain applicable
even in the presence of ground-state degeneracy.
Extensions of the present framework to degenerate or gapless regimes,
as well as to multi-particle sectors and dynamical response functions,
require additional analysis and remain an important direction for future investigation.

\section*{Acknowledgements}
The authors thank Hiroshi Ueda for bringing Ref.~\cite{CederbaumSM} to their attention.
The authors are grateful to Kai Phillip Schmidt for providing numerical data from Ref.~\cite{YangS2011} and
useful comments.
They also acknowledge stimulating discussions with Hiroshi Ueda, Shingo Kobayashi, Moritz Hirschmann,
Nic Shannon, and Shintaro Hoshino.

\paragraph{Funding information}
This work was supported by KAKENHI Grant No.\ JP20K03778 from the Japan Society for the Promotion of Science
(JSPS).

\begin{appendix}
\numberwithin{equation}{section}

\section{Proofs of Theorems~\ref{thm:1} and \ref{thm:2}}\label{sec:proof}

This appendix provides rigorous proofs of Theorem~\ref{thm:1} in Sec.~\ref{sec:BD_eff_H},
and Theorem~\ref{thm:2} in Sec.~\ref{sec:min_def_cluster_add}.
The key technical ingredient is a standard solution of the unitary Procrustes problem,
which we formulate as the following lemma.

\begin{lemma}
Let $\bm{A} \in \mathbb{C}^{n\times n}$ be invertible. Then the minimizer of
\[
\min_{\bm{U}\in U(n)} \|\bm{A}\bm{U}-\bm{1}_n\|_F
\]
is uniquely given by
\[
\bm{U}_{\mathrm{opt}} = (\bm{A}^\dagger \bm{A})^{-1/2}\bm{A}^\dagger.
\]
Equivalently, if
$\bm{A}=\bm{W}\bm{\Sigma}\bm{V}^\dagger$
is a singular-value decomposition (SVD) of $\bm{A}$, then
\[
\bm{U}_{\mathrm{opt}}=\bm{V}\bm{W}^\dagger.
\]
\end{lemma}

\begin{proof}[Proof of Lemma 1]
We include a short proof for completeness.
Let $\bm{A}=\bm{W}\bm{\Sigma}\bm{V}^\dagger$ be a SVD of $\bm{A}$,
where ${\bm W},{\bm V}\in U(n)$ and
${\bm \Sigma}=\mathrm{diag}(\sigma_1,\dots,\sigma_n)$ with $\sigma_j > 0$.
For any ${\bm U}\in U(n)$, setting ${\bm X}:={\bm V}^\dagger {\bm U}{\bm W} \in U(n)$,
we obtain
\begin{align*}
\|{\bm A}{\bm U}-\mathbf{1}_n\|_F^2
&=
\|\bm{\Sigma} \bm{X}-\mathbf{1}_n\|_F^2 =
\mathrm{Tr}({\bm \Sigma}^2)-2\,\mathrm{Re}\,\mathrm{Tr}({\bm \Sigma}{\bm X})+n \\
&
=
\sum_{j=1}^n \sigma_j^2
-2\sum_{j=1}^n \sigma_j\,\mathrm{Re}\,X_{jj}
+n.
\end{align*}
Since $\bm{X}$ is unitary, one has $|X_{jj}| \le 1$.
Because $\sigma_j>0$ for all $j$, the minimum is attained only if
$\mathrm{Re}(X_{jj})=1$ for all $j$, which implies $X_{jj}=1$ for all $j$.
A unitary matrix with all diagonal entries equal to $1$ must be the identity,
hence $\bm{U}_{\mathrm{opt}}=\bm{V}\bm{W}^\dagger$.

Finally, we show that this expression is independent
of the choice of singular vectors.
Define
$\bm{Q}:=(\bm{A}^\dagger \bm{A})^{-1/2}\bm{A}^\dagger$.
Since $\bm{A}$ is invertible, $\bm{A}^\dagger \bm{A}$ is positive definite,
so $\bm{Q}$ is uniquely determined by $\bm{A}$.
Using the SVD of $\bm{A}$, one finds
$\bm{Q}
=
(\bm{A}^\dagger \bm{A})^{-1/2}\bm{A}^\dagger
=
\bm{V}\bm{W}^\dagger$.
Hence $\bm{U}_{\mathrm{opt}}=\bm{Q}$, which is uniquely determined by $\bm{A}$.
This proves uniqueness.
\end{proof}

Thus the expression $\bm{V}\bm{W}^\dagger$ is uniquely determined by $\bm{A}$,
even though $\bm{W}$ and $\bm{V}$ themselves are not unique when the singular values are degenerate.

\begin{proof}[\textbf{Proof of Theorem 1}]
As in Eq.~\eqref{eq:SF}, we write $\bm{T}=\bm{S}\bm{F}$, where $\bm{F}$ is block-diagonal unitary.
Then
$\bm{T}_{11} = \bm{S}_{11}\bm{F}_{11}$ with
$\bm{F}_{11}\in U(n)$.
By Criterion 1, $\bm{F}_{11}$ minimizes
\[
\|\bm{T}_{11}-\bm{1}_n\|_F
=
\|\bm{S}_{11}\bm{F}_{11}-\bm{1}_n\|_F.
\]
Since $\bm{S}_{11}$ is invertible by assumption, Lemma 1 applies and yields the unique minimizer
$\bm{F}_{11}=\bm{V}_1 \bm{U}_1^\dagger$, where we have used the SVD of $\bm{S}_{11}$  in Eq.~\eqref{eq:SVD}.
%
Hence
\[
\bm{T}_{11}
=
\bm{S}_{11}\bm{F}_{11}
=
\bm{U}_1 \bm{\Sigma}_1 \bm{U}_1^\dagger.
\]
Finally, since
$\bm{F}_{11}\bm{H}_{\mathrm{eff},11}\bm{F}_{11}^\dagger=\bm{\Lambda}_1$,
we obtain
\[
\bm{H}_{\mathrm{eff},11}
=
\bm{F}_{11}^\dagger \bm{\Lambda}_1 \bm{F}_{11}
=
\bm{U}_1 \bm{V}_1^\dagger
\bm{\Lambda}_1
\bm{V}_1 \bm{U}_1^\dagger.
\]
This completes the proof.
\end{proof}


\begin{proof}[\textbf{Proof of Theorem 2}]
The proof proceeds in the same way, applied to the block
$\tilde{\bm T}_{m+1,m+1}$ in a fixed $m$-excitation sector.
We write
$\tilde{\bm T}=\tilde{\bm S}\tilde{\bm F}$,
where $\tilde{\bm F}$ is restricted to be block-diagonal unitary.
Then
$\tilde{\bm T}_{m+1,m+1}
=
\tilde{\bm S}_{m+1,m+1}\tilde{\bm F}_{m+1,m+1}$ with
$\tilde{\bm F}_{m+1,m+1}\in U(n)$.
By Criterion 2, $\tilde{\bm F}_{m+1,m+1}$ minimizes
\[
\|\tilde{\bm T}_{m+1,m+1}-\bm{1}_n\|_F
=
\|\tilde{\bm S}_{m+1,m+1}\tilde{\bm F}_{m+1,m+1}-\bm{1}_n\|_F.
\]
Since $\tilde{\bm S}_{m+1,m+1}$ is invertible as assumed, Lemma 1 applies and yields the unique minimizer
$\tilde{\bm F}_{m+1,m+1}
=
\tilde{\bm V}_{m+1}
\tilde{\bm U}_{m+1}^\dagger$, where we have used the SVD of $\tilde{\bm S}_{m+1,m+1}$ in Eq.~\eqref{eq:SVD_blocks}.
Hence
\[
\tilde{\bm T}_{m+1,m+1}
=
\tilde{\bm U}_{m+1}
\tilde{\bm \Sigma}_{m+1}
\tilde{\bm U}_{m+1}^\dagger.
\]
Finally, since $\tilde{\bm F}_{m+1,m+1}$ is unitary, we obtain
\[
\tilde{\bm H}_{\mathrm{eff},m+1,m+1}
=
\tilde{\bm F}_{m+1,m+1}^\dagger
\bm{\Lambda}_{m+1,m+1}
\tilde{\bm F}_{m+1,m+1}
=
\tilde{\bm U}_{m+1}
\tilde{\bm V}_{m+1}^\dagger
\bm{\Lambda}_{m+1,m+1}
\tilde{\bm V}_{m+1}
\tilde{\bm U}_{m+1}^\dagger.
\]
This proves the theorem.
\end{proof}

\section{Sketch of the proof of cluster additivity for $m$-excitation sectors}
\label{sec:proof_cluster_add}
We sketch an inductive proof of cluster additivity in the $m$-excitation sector.
For disconnected clusters,
it is not immediately evident from Ref.~\cite{Hormann2023} that
the transformed block $\bar{\bm{S}}_{m+1,m+1}$ has the required block-diagonal structure.
Since the diagonal blocks $\bar{\bm{S}}_{m+1,m+1}$ and
$\tilde{\bm{S}}_{m+1,m+1}$ coincide, as shown in Sec.~\ref{subsec:similarity_basis_transformation},
we discuss the properties of $\tilde{\bm{S}}_{m+1,m+1}$ in the following.

The key step is the cancellation of mixed-sector contributions
in the Schur-complement construction.
We have verified this cancellation explicitly,
but omit the full computation for brevity.

\begin{proof}[Sketch of Proof]
{\it Base case ($m=1$).}
The case $m=1$ is discussed in Sec.~\ref{subsec:similarity_basis_transformation}.
From the explicit expression in Eq.~\eqref{eq:transformed_state_one},
both Eqs.~\eqref{eq:cluster_additive_goal}
and \eqref{eq:upper_triangular_goal} hold for $m=1$.

{\it Inductive hypothesis.}
Assume that for all integers $n$ with $1\le n \le m-1$, the transformed eigenvector matrices
$\tilde{\bm{S}}^{(n)}$ satisfy the two properties stated in
Eqs.~\eqref{eq:cluster_additive_goal} and \eqref{eq:upper_triangular_goal}.
In particular,
$\tilde{\bm{S}}^{(m-1)}$ is block upper triangular up to the $(m-1)$th column,
and its diagonal blocks up to $(m,m)$ are block-diagonal for disconnected clusters.

{\it Inductive step.}
We define the next transformed matrix by
\begin{equation}
\tilde{\bm{S}}^{(m)} = \bm{W}_{m}^{-1}\,\tilde{\bm{S}}^{(m-1)},
\end{equation}
where $\bm{W}_{m}$ is given by Eq.~\eqref{eq:Wn_explicit}.
Using the decomposition
\(
\mathcal{H}=\mathcal{H}^{(\le m-2)}\oplus\mathcal{H}^{(m-1)} \oplus\mathcal{H}^{(\ge m)}
\),
corresponding to the index partition $I_{m-1}\mid \{m\}\mid I_{m}^{\mathrm{c}}$,
the matrix $\tilde{\bm{S}}^{(m-1)}$ takes the form
\begin{equation}
\tilde{\bm{S}}^{(m-1)}
=
\begin{pmatrix}
\tilde{\bm{S}}^{(m-1)}_{I_{m-1},I_{m-1}} & \tilde{\bm{S}}^{(m-1)}_{I_{m-1},m} & \tilde{\bm{S}}^{(m-1)}_{I_{m-1},I_{m}^\mathrm{c}} \\[4pt]
\bm{0}                         & \tilde{\bm{S}}^{(m-1)}_{m,m} & \tilde{\bm{S}}^{(m-1)}_{m,I_{m}^\mathrm{c}} \\[4pt]
\bm{0}          & \tilde{\bm{S}}^{(m-1)}_{I_{m}^\mathrm{c},m} &
\tilde{\bm{S}}^{(m-1)}_{I_{m}^\mathrm{c},I_{m}^\mathrm{c}}
\end{pmatrix}.
\label{eq:Sm_block_form}
\end{equation}
Applying $\bm{W}_m^{-1}$ yields
\begin{equation}
\tilde{\bm{S}}^{(m)}
=\bm{W}_{m}^{-1}\,\tilde{\bm{S}}^{(m-1)}=
\begin{pmatrix}
\tilde{\bm{S}}^{(m-1)}_{I_{m-1},I_{m-1}} & \tilde{\bm{S}}^{(m-1)}_{I_{m-1},m} & \tilde{\bm{S}}^{(m-1)}_{I_{m-1},I_{m}^\mathrm{c}} \\[4pt]
\bm{0}                         & \tilde{\bm{S}}^{(m-1)}_{m,m} & \tilde{\bm{S}}^{(m-1)}_{m,I_{m}^\mathrm{c}} \\[4pt]
\bm{0}          & \bm{0} &
\tilde{\bm{S}}^{(m)}_{I_{m}^\mathrm{c},I_{m}^\mathrm{c}}
\end{pmatrix},
\end{equation}
where
$(I_{m}^{\mathrm{c}},I_{m}^{\mathrm{c}})$ block is given by the Schur complement
\begin{equation}
\tilde{\bm{S}}^{(m)}_{I_{m}^{\mathrm c},I_{m}^{\mathrm c}}
=
\tilde{\bm{S}}^{(m-1)}_{I_{m}^{\mathrm c},I_{m}^{\mathrm c}}
-
\tilde{\bm{S}}^{(m-1)}_{I_{m}^{\mathrm c},m}
\bigl(
\tilde{\bm{S}}^{(m-1)}_{m,m}
\bigr)^{-1}
\tilde{\bm{S}}^{(m-1)}_{m,I_{m}^{\mathrm c}}.
\end{equation}
Thus, $\tilde{\bm{S}}^{(m)}$ is block upper triangular up to the $m$th column.

{\it Verification of the inductive properties.}
Equation~\eqref{eq:upper_triangular_goal} follows immediately from the construction above.
To verify Eq.~\eqref{eq:cluster_additive_goal},
consider the decomposition
\begin{equation}
\mathcal{H}^{(m)}_{A\cup B}
=
\bigoplus_{p=0}^{m}
\left(\mathcal{H}^{(m-p)}_A\otimes\mathcal{H}^{(p)}_B \right).
\end{equation}
The inductive hypothesis of $\tilde{\bm{S}}^{(m-1)}$
simplifies the evaluation of
\begin{equation}
\tilde{\bm{S}}^{(m)\,A\cup B}_{m+1,m+1}
=
\tilde{\bm{S}}^{(m-1)\,A\cup B}_{m+1,m+1}
-
\tilde{\bm{S}}^{(m-1)\,A\cup B}_{m+1,m}
\bigl(
\tilde{\bm{S}}^{(m-1)\,A\cup B}_{m,m}
\bigr)^{-1}
\tilde{\bm{S}}^{(m-1)\,A\cup B}_{m,m+1}.
\end{equation}
The Schur-complement expression contains terms that mix different values of $p$,
but these contributions cancel,
yielding a block-diagonal structure with respect to this decomposition:
\begin{equation}
\tilde{\bm{S}}^{(m)\,A\cup B}_{m+1,m+1}
=
\bigoplus_{p=0}^{m}
\left(
\tilde{\bm{S}}^{(m)\,A}_{m+1-p,m+1-p}
\otimes
\tilde{\bm{S}}^{(m)\,B}_{p+1,p+1}
\right).
\end{equation}
This establishes Eq.~\eqref{eq:cluster_additive_goal} at step $m$ and
completes the proof.
\end{proof}

\section{Comparison of the minimal transformation and the continuous unitary transformation in block diagonalization}\label{sec:comp}
Two effective Hamiltonians derived using different unitary transformations
are related through an additional unitary transformation.
This appendix compares two approaches: the CSM transformation and the continuous unitary
transformation.
We analyze the effective Hamiltonians for the one-dimensional transverse-field Ising model
discussed
in Sec.~\ref{sec:1d} and compare our results with those obtained using the continuous unitary transformation
in Ref.~\cite{YangS2011}.

\begin{figure}
  \centering
  \includegraphics[width=8.5cm]{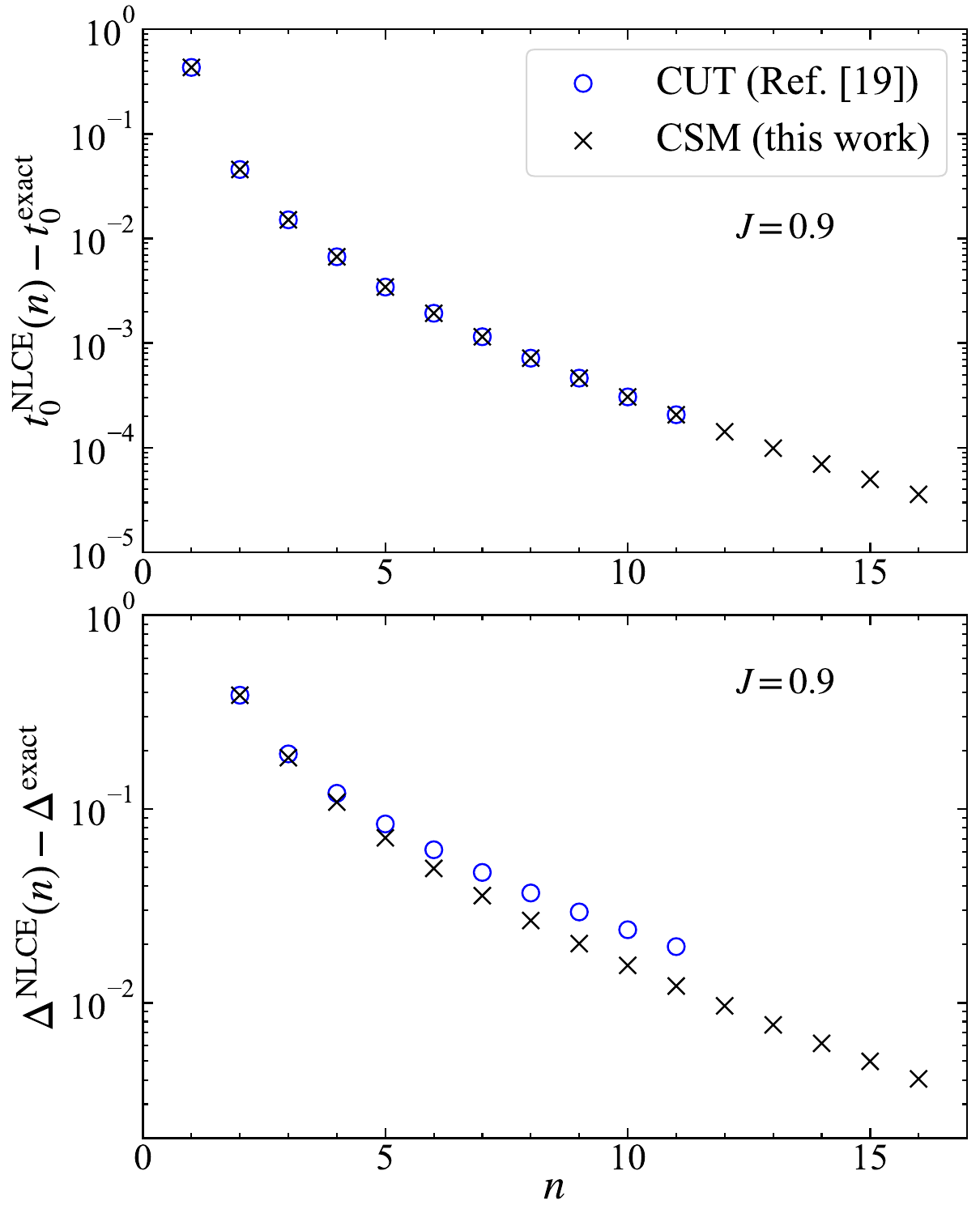}
  \caption{Comparison of the numerical linked-cluster expansion (NLCE) results obtained using
  the CSM transformation and the continuous unitary transformation (CUT)
  in block-diagonalization. Differences between the NLCE results and exact values are plotted.
  (Upper) On-site potential $t_0^{\rm NLCE}(n)$.
  (Lower) Energy gap $\Delta^{\rm NLCE}(n)$.
  Data for the CUT are adapted from Ref.~\cite{YangS2011}.
  }\label{fig:comp}
\end{figure}

The sum of the on-site potential, $\sum_i t_{ii}(n)$, corresponds to the trace of
the effective Hamiltonian,
$\sum_{i=1}^n t_{ii}(n) = \mathrm{Tr}{\bm H}_\text{eff,11}$.
This quantity is invariant under unitary transformations.
As a result, the on-site potential $t_{0}^{\rm NLCE}(n)$, derived from NLCE of $t_{ii}(n)$ up to $n$ spins,
is independent of the specific block-diagonalization method employed.
This invariance is confirmed by the data shown in Fig.~\ref{fig:comp}(Upper).

In contrast, the hopping amplitudes $t_{r}^{\rm NLCE}(n)$, extracted from NLCE of the off-diagonal
components $[{\bm H}_\text{eff,11}]_{i,i+r}$, do depend on the choice of transformation.
Figure~\ref{fig:comp}(Lower) shows that the CSM transformation yields faster convergence of the energy
gap,
\begin{equation}
\Delta^{\rm NLCE}(n)=t_0^{\rm NLCE}(n)+\sum_{r=1}^{n} t_{r}^{\rm NLCE}(n),
\end{equation}
to the exact value compared to the continuous unitary transformation.

\section{Symmetry analysis of the triplon bands in SrCu$_2$(BO$_3$)$_2$}
\subsection{Symmetry constraints on the triplon hopping matrices}\label{sec:sym_const}

SrCu$_2$(BO$_3$)$_2$ exhibits $S_4$ symmetry around the center of each square formed by four dimers, as well as
$C_{2V}$ symmetry around the center of each dimer.
The $S_4$ operation consists of a $\pi/2$ rotation around the $z$-axis combined with a mirror reflection in the
$xy$-plane. The $C_{2V}$ symmetry includes a $\pi$-rotation around the $z$-axis $C_2(z)$, a mirror reflection
in the $xz$-plane $\sigma_{xz}$, and a mirror reflection in the $yz$-plane $\sigma_{yz}$.
In a magnetic field applied along the $z$ axis, the system retains symmetry under the following operations:
(1) $S_4$ around the center of each square composed of four adjacent dimers;
(2) $C_2(z)$, $U_T \Theta \sigma_{xz}$, and $U_T \Theta \sigma_{yz}$ about the center of
  each dimer.
Here,
$U_T=\otimes_{{\bm r}\in\Lambda,m=1,2} (-i \sigma^x_{{\bm r},m})$ and $\Theta$ denotes complex conjugation.

The operations $U_T \Theta \sigma_{yz}$ and $C_2(z)$ transform the triplon operators as
${\bm t}_{\bm r} \rightarrow R_{\sigma(x)}{\bm t}_{\bm r}$ and ${\bm t}_{\bm r} \rightarrow R_{\pi(z)}{\bm
t}_{\bm r}$, respectively, for ${\bm r}\in\Lambda_A$,
with
\begin{align}
R_{\sigma(x)} =      \begin{bmatrix}
                       -1 & 0 & 0 \\
                       0 & 1 & 0 \\
                       0 & 0 & 1 \\
                     \end{bmatrix},\quad\quad
R_{\pi(z)} =         \begin{bmatrix}
                       1 & 0 & 0 \\
                       0 & 1 & 0 \\
                       0 & 0 & -1 \\
                     \end{bmatrix}.
\end{align}
The $S_4$ operation transforms
${\bm t}_{\bm r} \rightarrow S_{\alpha\rightarrow \bar{\alpha}} {\bm t}_{{\bm r}'}$ for ${\bm
r}\in\Lambda_\alpha$ and ${\bm r}'\in\Lambda_{\bar{\alpha}}$,
with
\begin{align}
S_{A\rightarrow B} = \begin{bmatrix}
                       0 & 1 & 0 \\
                       -1 & 0 & 0 \\
                       0 & 0 & -1 \\
                     \end{bmatrix},\quad\quad
S_{B\rightarrow A} = \begin{bmatrix}
                       0 & -1 & 0 \\
                       1 & 0 & 0 \\
                       0 & 0 & 1 \\
                     \end{bmatrix}.
\end{align}

The on-site potential ${{\bm t}^{\dagger}_{{\bm r}}} {\bm M}_{AA,(0,0)} {\bm t}_{{\bm r}}$
  (${\bm r}\in\Lambda_A$) is invariant under $C_2(z)$, $U_T \Theta \sigma_{xz}$, and $U_T \Theta \sigma_{yz}$.
The invariance conditions under $C_2(z)$ and $U_T \Theta \sigma_{yz}$ are given by
$ R_{\pi(z)}^\dagger {\bm M}_{AA,(0,0)}R_{\pi(z)} = {\bm M}_{AA,(0,0)}$
and $ R_{\sigma(x)}^\dagger {\bm M}^\ast_{AA,(0,0)}R_{\sigma(x)} = {\bm M}_{AA,(0,0)}$,
which result in the matrix form given in Eq.~(\ref{eq:M00M01}).
This matrix also
satisfies $U_T \Theta \sigma_{xz}$ invariance.
By $S_4$ symmetry, the on-site potential for $B$-type dimers becomes
\begin{align}
{\bm M}_{BB,(0,0)}=  \begin{bmatrix}
                       R_{02} & I_{01} & 0 \\
                      -I_{01} & R_{01} & 0 \\
                       0 & 0 & R_{03} \\
                     \end{bmatrix}.
\end{align}
All $R_i$ and $I_i$ in this appendix denote real and imaginary numbers, respectively.

For the nearest-neighbor hopping term
${{\bm t}^{\dagger}_{{\bm r}+(1,0)}} {\bm M}_{BA,(1,0)} {\bm t}_{{\bm r}}$
(${\bm r}\in\Lambda_A$), $U_T \Theta \sigma_{xz}$ invariance imposes
Eq.~(\ref{eq:M00M01}).
From the $U_T \Theta \sigma_{yz}$ and $C_2(z)$ symmetry, we obtain
\begin{align}
{\bm M}_{BA,(-1,0)} =
            \begin{bmatrix}
              I_{11} & R_{11} & -I_{12} \\
              R_{12} & I_{13} & -R_{13} \\
              -I_{14} & -R_{14} & I_{15} \\
            \end{bmatrix}.
\end{align}
$S_4$ symmetry further leads to
\begin{align}
{\bm M}_{AB,(0,1)} &=
            \begin{bmatrix}
              -I_{13} &  R_{12} & -R_{13} \\
               R_{11} & -I_{11} &  I_{12} \\
              -R_{14} &  I_{14} & -I_{15} \\
            \end{bmatrix},\quad\quad
 {\bm M}_{AB,(0,-1)}
=           \begin{bmatrix}
              -I_{13} & R_{12} &  R_{13} \\
              R_{11} & -I_{11} & -I_{12} \\
              R_{14} & -I_{14} & -I_{15} \\
            \end{bmatrix}.
\end{align}

For second-neighbor hopping, ${\bm M}_{AA,(1,1)}$ and its Hermitian conjugate satisfy $C_2(z)$ symmetry,
which results in
\begin{align}
{\bm M}_{AA,(1,1)} =
            \begin{bmatrix}
              R_{21} & C_{21} & C_{22} \\
              C_{21}^\ast & R_{22} & C_{23} \\
              -C_{22}^\ast & -C_{23}^\ast & R_{23} \\
            \end{bmatrix},
\end{align}
where $C_i$ denote complex numbers. From the $U_T \Theta \sigma_{yz}$ invariance, we have
\begin{align}
{\bm M}_{AA,(-1,1)} =
            \begin{bmatrix}
              R_{21} & -C_{21}^\ast & -C_{22}^\ast \\
              -C_{21} & R_{22} & C_{23}^\ast \\
              C_{22} & -C_{23} & R_{23} \\
            \end{bmatrix}.
\end{align}
$S_4$ symmetry further provides
\begin{align}
{\bm M}_{BB,(-1,1)} &=
            \begin{bmatrix}
              R_{22} & -C_{21}^\ast & C_{23} \\
              -C_{21} & R_{21} & -C_{22} \\
              -C_{23}^\ast & C_{22}^\ast & R_{23} \\
            \end{bmatrix},\quad\quad
{\bm M}_{BB,(1,1)}  =
            \begin{bmatrix}
              R_{22} & C_{21} & -C_{23}^\ast \\
              C_{21}^\ast & R_{21} & -C_{22}^\ast \\
              C_{23} & C_{22} & R_{23} \\
            \end{bmatrix}.
 \end{align}

For third-neighbor hopping, ${\bm M}_{AA,(2,0)}$ is invariant under both
$U_T \Theta \sigma_{xz}$ and $C_2 (z)$, and $S_4$ symmetry relates this to ${\bm M}_{BB,(0,2)}$,
which results in
\begin{align}
{\bm M}_{AA,(2,0)} &=
            \begin{bmatrix}
              R_{31} & I_{31} & R_{32} \\
              -I_{31} & R_{33} & I_{32} \\
              -R_{32} & I_{32} & R_{34} \\
            \end{bmatrix},\quad\quad
{\bm M}_{BB,(0,2)} =
            \begin{bmatrix}
              R_{33} & I_{31} &  I_{32} \\
              -I_{31} & R_{31} & -R_{32} \\
               I_{32} & R_{32} & R_{34} \\
            \end{bmatrix}.
\end{align}
Another third-neighbor hopping matrix ${\bm M}_{AA,(0,2)}$ and its symmetry operation are given by
\begin{align}
{\bm M}_{AA,(0,2)} &=
            \begin{bmatrix}
              R'_{31} & I'_{31} & I'_{32} \\
              -I'_{31} & R'_{32} & R'_{33} \\
              I'_{32} & -R'_{33} & R'_{34} \\
            \end{bmatrix},\quad\quad
{\bm M}_{BB,(2,0)} =
            \begin{bmatrix}
              R'_{32} & I'_{31} & -R'_{33} \\
              -I'_{31} & R'_{31} & I'_{32} \\
               R'_{33} & I'_{32} & R'_{34} \\
            \end{bmatrix},
\end{align}
where $R'_i$ and $I'_i$ denote real and imaginary numbers, respectively.

\subsection{Symmetry-protected degeneracy in the triplon bands}\label{sec:Kramers}

This appendix discusses the symmetry origin of the band degeneracy at the
Brillouin-zone boundary.
The analysis serves as a consistency check of the effective Hamiltonian
constructed in the main text.

As discussed in Appendix~\ref{sec:sym_const},
SrCu$_2$(BO$_3$)$_2$ in a magnetic field applied along the $z$-axis
preserves the symmetries generated by $S_4$, $U_T\Theta\sigma_{xz}$, and $U_T\Theta\sigma_{yz}$.
The system is therefore also invariant under the combined antiunitary operations
$A_x:= U_T\Theta\sigma_{xz} S_4$ and $A_y:=U_T\Theta\sigma_{yz} S_4$.

These operations act on spatial coordinates and spin components as
\begin{align}
A_x:\quad &(x,y,z) \mapsto (-y,-x-1,-z), \qquad (S^x,S^y,S^z) \mapsto (-S^y,-S^x,S^z), \\
A_y:\quad &(x,y,z) \mapsto (y+1,x,-z), \qquad (S^x,S^y,S^z) \mapsto (S^y,S^x,S^z).
\end{align}
Here $A_x$ preserves the site labels 1 and 2 within each dimer, whereas $A_y$ interchanges them.
The spatial parts correspond to screw-type operations,
which generate lattice translations upon squaring.

\paragraph{Action in the one-triplon sector.}
We consider the triplon creation operators
$t_{\bm r}^{\mu\dagger}
= i S^\mu_{\bm{r},1} - i S^\mu_{\bm{r},2}$
$(\mu=x,y,z)$
defined on each dimer.
Denoting the spatial part of $A_x$ by $g_x(\bm{r})$, we obtain
\begin{align}
A_x\, t_{\bm{r}}^{x\dagger}\, A_x^{-1}
&= t_{g_x(\bm{r})}^{y\dagger}, \qquad
A_x\, t_{\bm{r}}^{y\dagger}\, A_x^{-1}
= t_{g_x(\bm{r})}^{x\dagger}, \qquad
A_x\, t_{\bm{r}}^{z\dagger}\, A_x^{-1}
= -\,t_{g_x(\bm{r})}^{z\dagger}.
\end{align}

The dimer-singlet vacuum $|s\rangle$, defined as the product of singlet states on each dimer,
is invariant under $A_x$ up to an overall phase,
since it is invariant under time reversal and under any common spin rotation
acting on the two spins within a dimer.
Therefore, the one-triplon states
$|t_\mu,\bm{r}\rangle = t_{\bm{r}}^{\mu\dagger}|s\rangle$
transform as
\begin{align}
A_x |t_x,\bm{r}\rangle &= |t_y, g_x(\bm{r})\rangle, \qquad
A_x |t_y,\bm{r}\rangle = |t_x, g_x(\bm{r})\rangle, \qquad
A_x |t_z,\bm{r}\rangle = -|t_z, g_x(\bm{r})\rangle.
\end{align}

Applying $A_x$ twice and using $g_x^2(\bm{r}) = \bm{r} + (1,-1)$, we obtain
\begin{equation}
A_x^2 |t_\mu,\bm{r}\rangle
= |t_\mu, \bm{r} + (1,-1)\rangle,
\end{equation}
which implies
\begin{equation}
A_x^2 = T_{(1,-1)}
\end{equation}
in the one-triplon sector.
Similarly, one finds from the corresponding transformation
\begin{equation}
A_y^2 = T_{(1,1)}.
\end{equation}

\paragraph{Kramers degeneracy at the Brillouin-zone boundary.}
For Bloch eigenstates with momentum $\bm k$, the translation operator yields
$T_{\bm r} \to e^{i \bm r \cdot \bm k}$.
Therefore,
$A_x^2 = e^{i(k_x-k_y)}$ and
$A_y^2 = e^{i(k_x+k_y)}$.
Along the Brillouin-zone boundaries defined by
$k_x - k_y = (2n+1)\pi$ or $k_x + k_y = (2n+1)\pi$,
these operators satisfy $A_{x,y}^2 = -1$.

At these momenta, the antiunitary symmetries $A_x$ or $A_y$
map $\bm k$ onto itself up to a reciprocal lattice vector.
Hence, for any Bloch eigenstate $|u_{\bm k}\rangle$,
the state $A_{x,y}|u_{\bm k}\rangle$ is orthogonal to $|u_{\bm k}\rangle$,
since $A_{x,y}$ is antiunitary and satisfies $A_{x,y}^2=-1$.
Moreover, it has the same energy, implying a symmetry-protected double degeneracy.

This degeneracy enforces band crossings of the triplon spectrum
along the Brillouin-zone boundary.
In the quasi-two-dimensional structure of SrCu$_2$(BO$_3$)$_2$,
these degeneracies extend to nodal planes in three-dimensional momentum space.
Similar Kramers degeneracies
and nodal planes in spin excitations have been reported in a spin model for Volborthite~\cite{Furukawa2020}
and in the Kitaev--Heisenberg model~\cite{Corticelli2022}.

\end{appendix}

\bibliography{note}

\end{document}